\documentclass[12pt]{article}
\usepackage{latexsym}
\usepackage{epsfig,amssymb,euscript}
\usepackage{amsmath}
\numberwithin{equation}{section}  
\newsavebox{\ns}
\newsavebox{\dbrane}
\newsavebox{\dbshort}

\def\be{\begin{equation}}
\def\ee{\end{equation}}
\def\bea{\begin{eqnarray}}
\def\eea{\end{eqnarray}}

\def\be{\begin{equation}}
\def\ee{\end{equation}}
\def\ba{\begin{eqnarray}}
\def\ea{\end{eqnarray}}
\newcommand{\per}{\ell}
\newcommand{\nn}{\nonumber}

\def\Dslash{\,\,{\raise.15ex\hbox{/}\mkern-12mu D}}
\def\Dbarslash{\,\,{\raise.15ex\hbox{/}\mkern-12mu {\bar D}}}
\def\delslash{\,\,{\raise.15ex\hbox{/}\mkern-9mu \partial}}
\def\delbarslash{\,\,{\raise.15ex\hbox{/}\mkern-9mu {\bar\partial}}}
\def\pslash{\,\,{\raise.15ex\hbox{/}\mkern-9mu p}}
\def\calDslash{\,\,{\raise.15ex\hbox{/}\mkern-12mu {\cal D}}}

\newcommand\R{\mathbb{R}}
\newcommand\Z{\mathbb{Z}}
\newcommand\F{\mathbb{F}}

\newcommand\cp{\mathbb{CP}}
\newcommand\C{\mathbb{C}}
\newcommand\T{\mathbb{T}}
\newcommand\diff{\mathrm{d}}
\newcommand{\de}{\partial}

\begin{document}
\begin{titlepage}
\begin{center}
\today
{\small\hfill hep-th/0411238}\\
{\small\hfill CERN-PH-TH/2004-222}\\
{\small\hfill HUTP-04/A0046}\\

\vskip 1.3cm 
{\Large \bf Toric Geometry, Sasaki--Einstein Manifolds} \\ 
\vskip 4mm
{\Large \bf and a New Infinite Class of AdS/CFT Duals}
\vskip 8mm
{Dario Martelli$^{1}$~ and ~ James Sparks$^{2}$}\\
\vskip 5mm

{\sl 1: Department of Physics, CERN Theory Division\\
1211 Geneva 23, Switzerland\\
\vskip 4mm
2: Department of Mathematics, Harvard University \\
One Oxford Street, Cambridge, MA 02318, U.S.A.\\
{\it and} \\
Jefferson Physical Laboratory, Harvard University \\
Cambridge, MA 02138, U.S.A.\\
\vskip 4mm
 {\tt dario.martelli@cern.ch} \ \ {\tt
sparks@math.harvard.edu}\\}

\end{center}

\vskip 5mm 
\begin{abstract}

\vskip 3mm
\noindent
Recently an infinite family of explicit Sasaki--Einstein metrics
$Y^{p,q}$ on $S^2\times S^3$ has been discovered, 
where $p$ and $q$ are two coprime positive integers, with $q<p$. 
These give rise to a corresponding family of Calabi--Yau cones, which moreover are
{\em toric}.
Aided by several recent results in toric geometry,
we show that these are K\"ahler quotients ${\C^4//U(1)}$, namely the vacua of gauged linear
sigma models with charges $(p,p,-p+q,-p-q)$, thereby generalising the conifold,
which is $p=1,q=0$.
We present the corresponding toric diagrams and show that these may be embedded
in the toric diagram for the orbifold  $\C^3/\Z_{p+1}\times\Z_{p+1}$
for all $q<p$ with fixed $p$. We hence find that the $Y^{p,q}$ manifolds
are AdS/CFT dual to an infinite class of ${\cal N}=1$ superconformal field
theories arising as IR fixed points of toric quiver gauge 
theories with gauge group $SU(N)^{2p}$. As a
non--trivial example, we show that $Y^{2,1}$ is an explicit 
\emph{irregular} Sasaki--Einstein metric
on the horizon of the complex cone over the first del Pezzo surface.
The dual quiver gauge theory has already been constructed for this case and
hence we can predict the exact central charge of this theory 
at its IR fixed point using the 
AdS/CFT correspondence. The value we obtain is a quadratic irrational
number and, remarkably, agrees with a recent purely field theoretic 
calculation using $a$--maximisation. 

\end{abstract}

\end{titlepage}
\pagestyle{plain}
\setcounter{page}{1}
\newcounter{bean}
\baselineskip18pt


\section{Introduction and summary}

The AdS/CFT correspondence \cite{Maldacena} predicts that type IIB string theory
on $AdS_5\times Y_5$, with appropriately chosen self--dual
five--form flux, is dual to an $\mathcal{N}=1$ four--dimensional
superconformal field theory whenever $Y_5$ is \emph{Sasaki--Einstein}
\cite{Kehagias,KW,acharya,MP}.
This latter condition may be defined as saying that the metric
cone over $Y_5$
\be
\diff s^2 (C(Y_5)) = \diff r^2 + r^2 \diff s^2 (Y_5)\label{metcone}\ee
is Ricci--flat K\"ahler {\it i.e.} Calabi--Yau. The superconformal
field theory may be thought of as arising from a stack of D3--branes
sitting at the tip of the Calabi--Yau cone. Notice that, unless
$Y_5$ is the round metric on $S^5$, appropriately normalised, the tip
of the cone at $r=0$ will be singular.

It is a striking fact that, until very recently, the only
Sasaki--Einstein five--manifolds that were known explicitly
in the literature\footnote{E. Calabi has constructed an
explicit K\"ahler--Einstein metric on del Pezzo $6$ -- recall that this
is the blow--up of $\cp^2$ at $6$ points -- with a certain symmetric
configuration of the $6$ blown--up points. The corresponding Sasaki--Einstein
metric on $\#6(S^2\times S^3)$ is thus also explicit. This metric has
apparently never been published. We thank S.--T. Yau for pointing this
out to us.}
were precisely the round metric on $S^5$ and the
homogeneous metric $T^{1,1}$ on $S^2\times S^3$, or quotients thereof.
For the five--sphere the Calabi--Yau cone is simply
$\C^3$ and the dual superconformal field theory is the maximally
supersymmetric $\mathcal{N}=4$ $SU(N)$ theory. For $T^{1,1}$
the Calabi--Yau cone is the conifold and the
dual $\mathcal{N}=1$ superconformal field theory was given in \cite{KW, MP}.

Due to the rather limited number of examples in the literature
detailed tests of the AdS/CFT conjecture for more interesting geometries have
been lacking\footnote{Although one can still deduce some geometric information
for the regular Sasaki--Einstein manifolds $\#l(S^2 \times S^3)$, 
which are $U(1)$ bundles over del Pezzo surfaces with $l$ points blown up, 
$l=3,\ldots,8$, even though the general metrics are not known explicitly.}. 
Indeed, one is restricted
to quotients (orbifolds) of $S^5$ and $T^{1,1}$. These have been extensively studied
using orbifold techniques which by now are completely standard.
For example, Klebanov and Witten argued that the
field theory for $T^{1,1}$ may be obtained via a relevant deformation of
the $\mathcal{N}=2$ orbifold $S^5/\Z_2$.

However, this has changed drastically 
with the recent discovery \cite{paper2} of a countably infinite class of \emph{explicit}
Sasaki--Einstein metrics on $Y^{p,q}\cong S^2\times S^3$.
These were initially found by reduction and T--duality of a class 
of supersymmetric M--theory solutions discovered in \cite{paper1}. 
The  family is characterised by two relatively
prime positive integers $p,q$, with $q<p$.
A particularly interesting feature of these Sasaki--Einstein manifolds is
that there are countably infinite classes which are both
\emph{quasi--regular} and \emph{irregular}. These
terms are not to be confused with regularity of the metric: the metrics
are all smooth metrics on $S^2\times S^3$. Rather, they refer to properties
of the orbits of a certain Killing vector field.
Indeed, on any Sasaki--Einstein manifold $Y$ there exists a canonically
defined Killing vector field $K$, called the \emph{Reeb} vector in the
mathematics literature. The orbits of this Killing vector field
may or may not close. If they close then there is a (locally free)
 $U(1)$ action on $Y$ and such Sasaki--Einstein manifolds are
called quasi--regular. The geometries $Y^{p,q}$ with
$4p^2-3q^2$ a square are examples of such manifolds.
If the orbits of the Reeb vector field do not close the Killing vector generates
an action of $\R$ on $Y$, with the orbits densely filling the orbits of
a torus, and the Sasaki--Einstein manifold
is said to be irregular. The geometries $Y^{p,q}$ with
$4p^2-3q^2$ \emph{not} a square are the first
examples of such geometries in the literature\footnote{Thus disproving a conjecture of Cheeger and Tian 
\cite{conjecture} that such examples do not exist. We thank the referee for drawing our attention to this reference.}. Another interesting 
feature of these metrics is that the volumes are always 
given by a \emph{quadratic irrational} number 
times the volume of the round metric on 
$S^5$ -- recall a quadratic irrational is of the form $a+b\sqrt{c}$ 
where $a,b\in \mathbb{Q}, c\in \mathbb{N}$. Moreover, the volumes are rationally 
related to that of $S^5$ if and only if the Sasaki--Einstein is quasi--regular. 

Recall that all four--dimensional $\mathcal{N}=1$
superconformal field theories possess an R--symmetry, commonly referred to as
the $U(1)$ R--symmetry. However,
crucially this symmetry is not always a $U(1)$ symmetry -- this is
true only if the R--charges of all the fields are \emph{rational}.
In general, this is not true, as exemplified by the recent work
of \cite{IW}. In the latter reference it is shown that the
exact R--symmetry of a superconformal field theory maximises a
certain combination of 't Hooft anomalies 
$a_{trial} (R)=(9\mathrm{Tr} R^3 - 3 \mathrm{Tr} R)/32$. 
The maximal value is then precisely the exact $a$ central charge of the 
superconformal field theory. Since one is maximising a cubic with rational coefficients, the
resulting R--charges are always algebraic numbers.
Recall that in  AdS/CFT the R--symmetry is precisely dual to the canonical 
Killing vector field $K$ discussed above. Moreover, the central charge $a_Y$ 
for the field theory dual to $Y$ is inversely proportional to its volume. 
In particular, we have \cite{gubser}
\be
\frac{a_Y}{a_{S^5}} = \frac{\mathrm{vol}(S^5)}{\mathrm{vol}(Y)}~.\label{central}\ee

It is thus clearly of interest to identify the dual superconformal field
theories for the Sasaki--Einstein manifolds $Y^{p,q}$, so as to compare the 
exact results on both sides of the duality.  In this paper we take the first 
substantial steps in this program
by analysing in considerable detail the geometry of the manifolds $Y^{p,q}$,
and the associated Calabi--Yau cones. The results allow us to show that the
metrics $Y^{p,q}$ are dual to a class of ${\cal N}=1$ 
superconformal field theories arising as IR fixed 
points of certain toric quiver gauge theories, with gauge group $SU(N)^{2p}$.

The case $p=2,q=1$  is somewhat special. This corresponds to the geometry
with largest volume, and is an irregular metric. The dual field theory 
therefore has the smallest central charge within the family, and moreover 
is expected to be {\em quadratic irrational}. Rather surprisingly, 
we find that the metric $Y^{2,1}$ turns out to be an explicit metric on the horizon
of the complex cone over the first del Pezzo surface. For this, the corresponding
$SU(N)^4$ quiver gauge theory and superpotential have already been identified \cite{hanany}.
We can then compute the central charge (\ref{central}) and also the R--charges 
of the baryons for this theory using AdS/CFT, where the baryons
correspond to D3--branes wrapped over 3--cycles whose metric cones are 
supersymmetric cycles (complex divisors) in the cone over $Y^{2,1}$. 
The values we find are all quadratic irrational numbers. At first sight
these results present a puzzle, as the central charge computed in 
\cite{IW,herzog2,herzog1} was found to be a rational number. 
However, a closer inspection of the 
quiver theory shows that the $a$--maximisation calculation is somewhat 
more subtle in this case\footnote{We are very grateful to M. Bertolini, 
F. Bigazzi, A. Hanany, K. Intriligator, and B. Wecht for discussions 
on this issue.}. 
Indeed, using $a$--maximisation \cite{IW} applied to the quiver theory, 
the authors of \cite{BB} find a central charge, as well as R--charges, 
which agree perfectly with the values obtained using the geometrical 
results of this paper.
This constitutes an extremely beautiful test of the AdS/CFT correspondence, as well 
as the general $a$--maximisation procedure advocated in \cite{IW}.

Given the results presented here, 
in principle the duals to the remaining geometries, 
with general $p$ and $q$, $q<p$, can be constructed using the ``toric
algorithm'' of \cite{hanany}. These will provide an infinite series 
of ${\cal N}=1$ superconformal field theories, whose central charges are
generically quadratic irrational. 
It will be interesting to obtain these explicitly, and to 
compare the results of $a$--maximisation for these theories with the various
geometrical results presented in this paper. However, we leave these 
calculations for future work.

As a final point, we note that in \cite{paper3} a generalisation of 
the metrics $Y^{p,q}$ to all dimensions was presented (see 
also references \cite{strasbourg} and \cite{morese} for a generalisation of this 
generalisation). In particular 
there are countably infinite classes of supersymmetric solutions 
$AdS_4\times Y_7$ to M--theory, which will have three--dimensional 
CFT duals, where the metric $Y_7$ is built using any positive curvature 
K\"ahler--Einstein metric in real dimension four \cite{paper3}. These have been 
classified \cite{tian, tianyau}. For the case when the K\"ahler--Einstein manifold is 
\emph{toric}, one has only three cases: $\cp^2$, $\cp^1\times\cp^1$, and  
$dP_3$, where the latter is the third del Pezzo surface. 
Using the techniques developed in this
paper, one can show that for the first two cases the metric cones over $Y_7$ are 
given by K\"ahler quotients $\C^5//U(1)$, and $\C^6//U(1)^2$, respectively, where 
the various $U(1)$ charges are, with appropriate definitions\footnote{In 
particular, the definitions here are different from those in \cite{paper3}.} 
of the Chern numbers
$p$ and $k$, $Q=(p,p,p,-3p+k,-k)$ and $Q_1=(p,p,0,0,-2p+k,-k)$, 
$Q_2=(0,0,p,p,-2p+k,-k)$, respectively.

\subsection*{Outline}

The first point to note about the manifolds $Y^{p,q}$, and their
associated Calabi--Yau cones, is that they are all \emph{toric}.
This essentially means that there is an effective action of a torus
$\T^3\cong U(1)^3$ on $C(Y^{p,q})$ which preserves the symplectic form of the cone
and commutes with the homothetic $\R^+$ action. 
Indeed, this torus action is an isometry, and so also preserves the
metric. The torus action and symplectic form then allow us to define
a \emph{moment map}, $\mu:C(Y^{p,q})\rightarrow \R^3$. The image
in $\R^3$ is always a \emph{good convex rational polyhedral cone} in
$\R^3$ \cite{L}. These terms will be explained more carefully later.
However, roughly this is a convex cone formed by intersecting some number
of planes through the origin. The moment map exhibits $C(Y^{p,q})$ as
a $\T^3$ fibration over this \emph{moment cone}, with the fibres 
collapsing over the faces, or \emph{facets}, of the cone in a 
way determined by the normal vectors to the facets. 
We shall find explicitly that the moment cone for $Y^{p,q}$
is a four--faceted good strictly convex rational polyhedral cone.

Having computed the moment cone for $C(Y^{p,q})$ we may then apply
a \emph{Delzant theorem} \cite{D} for symplectic toric cones worked
out recently in \cite{L}. In physics terms, this takes the combinatorial data
defining the moment cone and uses it to produce a \emph{gauged linear
sigma model} \cite{wittenLSM}. By construction the classical vacuum of the
linear sigma model is precisely the Calabi--Yau cone one started with.
More mathematically, this would be called a symplectic -- or, more
precisely, K\"ahler -- quotient of $\C^d$ by a compact abelian group.
The final result is:

\begin{itemize}
\item {\sl The metric cones over $Y^{p,q}$ are explicit Calabi--Yau
 metrics for 
the $U(1)$ gauged linear sigma model on $\C^4$ with charges $(p,p,-p+q,-p-q)$,
and zero Fayet--Iliopoulos parameter.}
\end{itemize}

If we denote the vacuum of a linear sigma model by $X=\C^4//U(1)$, then it is easy to see
that, rather generally, $c_1(X)=0$ is equivalent to the charges of the
$U(1)$ gauge group summing to zero. Clearly this is true for the gauged linear
sigma model above, and hence $X$ is indeed topologically Calabi--Yau.
In this process we lose precise information about the metric
-- in particular, the induced metric from $\C^4$ is \emph{not} Ricci--flat.
However, we have now gained an explicit description of the Calabi--Yau singularity.
Indeed, by constructing invariant monomials one also obtains an
algebraic description of the singularity. 

We may then give the \emph{toric diagram} for the Calabi--Yau
singularity. This may be realised as an integral polytope in $\R^2$.
Roughly, the four outward pointing primitive normal vectors that define
the moment cone
lie in a plane as a result of the Calabi--Yau condition. Projecting
these vectors onto this plane yields the vertices of the
toric diagram for a minimal presentation of the singularity.
We show that the resulting toric diagrams may all be
embedded inside that of the orbifold $\C^3/\Z_{p+1}\times\Z_{p+1}$
where the two factors are generated by $(\omega_{p+1},\omega_{p+1}^{-1},1)$,
$(\omega_{p+1},1,\omega_{p+1}^{-1})\subset SU(3)$, respectively, where
$\omega_{p+1}$ is
a $(p+1)$--th root of unity. The vertices of the polytope are then
$(0,0)$, $(0,p+1)$ and $(p+1,0)$ (the position of the origin is irrelevant)
and we show that the toric
diagram for $C(Y^{p,q})$ lives inside this polytope for all $q<p$ and
fixed $p$. Geometrically, this means that the Calabi--Yau cone
$C(Y^{p,q})$ may be obtained by (partial) toric crepant resolution
of the orbifold \cite{MP,Beasley:1999uz}.

Also, as part of our general analysis, we find a class of 
supersymmetric submanifolds in the geometries $C(Y^{p,q})$. 
Specifically, we show that the cones over the special orbits of 
the cohomogeneity one action on $Y^{p,q}$ are calibrated submanifolds -- 
in fact complex divisors -- of the Calabi--Yau. Recall that 
D3--branes wrapped over the horizon 3--cycles are 
dual to baryons in the AdS/CFT correspondence \cite{wittenbaryon,kleb}. 
We compute the volumes of these submanifolds, and hence give a prediction for 
the R--charges of the corresponding baryons.

Given the toric diagram for $C(Y^{p,q})$ there are methods to
construct a superconformal field theory, whose Higgs branch is 
the toric variety $X\cong C(Y^{p,q})$, purely from the
combinatorial data that defines $X$ \cite{hanany}. Indeed, the 
point is that the
field theory for the orbifold $\C^3/\Z_{p+1}\times\Z_{p+1}$,
in which the geometries are
``embedded'', is known from standard orbifold techniques.
The Calabi--Yau cones $C(Y^{p,q})$ are obtained by partial
resolution, which amounts to turning on specific combinations of
Fayet--Iliopoulos parameters in the gauged linear sigma model.
The field theories in question are then
rather conventional quiver gauge theories with polynomial superpotentials. 
The number of nodes of the quiver is simply twice the area 
of the toric diagram, which is $2p$ for all $q$ with fixed $p$.

Rather surprisingly, we find that the toric
diagram for $Y^{2,1}$ is precisely the same as that for the complex
cone over the first del Pezzo surface. Recall that the latter is
the blow--up of $\cp^2$ at one point, and that the complex cone 
over this is indeed a real cone over $S^2 \times S^3$. It follows that $Y^{2,1}$, which
is \emph{irregular}, is an explicit Sasaki--Einstein metric on the
horizon, or boundary, of this cone. This is interesting, since
the higher del Pezzo surfaces, which are $\cp^2$
with $3\leq r\leq 8$ generic points blown up, admit K\"ahler--Einstein
metrics \cite{tian, tianyau}. The complex cones then carry regular
Sasaki--Einstein metrics.
The case of one or two points blown up has always been something of a
puzzle, since these del Pezzos do not admit K\"ahler--Einstein metrics and thus the
Sasaki--Einstein metrics associated to the complex cones could not possibly be regular.
We have thus resolved this puzzle, at least in the case of one blow--up.

The quiver gauge theory dual to the complex cone over the first del Pezzo
surface has been presented in the literature \cite{hanany}. 
The AdS/CFT correspondence then predicts the exact central charge 
of this theory in the IR. Using the explicit metric $Y^{2,1}$,  
the result we obtain is 
\be
\frac{a_{S^5}}{a_{Y^{2,1}}}=\frac{\mathrm{vol}(Y^{2,1})}{\mathrm{vol}(S^5)} = \frac{13\sqrt{13}+46}{12\cdot 27}\sim 
\frac{7.74}{27}~.\label{ratio}\ee
Remarkably, this value coincides precisely with a recent application of 
$a$--maximisation \cite{IW} to the quiver gauge theory \cite{BB}.
Moreover, we also find perfect agreement for the charges of
($SU(2)_F$ singlet) baryons in the  gauge theory.

\vskip 4mm

The plan of the rest of the paper is as follows. In Section 2, after recalling 
some basic facts about Sasaki--Einstein geometry, we give a summary of the
construction of the metrics $Y^{p,q}$, and recall several of their features. 
Section 3 contains a review of symplectic toric geometry -- 
in particular toric contact geometry -- which we use extensively in the 
remainder of the paper. 
In Section 4 we compute the image of the moment map associated to 
the toric Calabi--Yau cones $C(Y^{p,q})$. In section 5 we apply a Delzant construction
to obtain a gauged liner sigma model (GLSM) description of the Calabi--Yau spaces.
Moreover we analyse directly the structure of the moduli space of vacua 
of the GLSM in Section 5.3. In Section 6 the associated toric Gorenstein
singularities are described. In Section 7 we demonstrate that $Y^{2,1}$
is an irregular metric on the horizon of the complex cone over the first 
del Pezzo surface, and exhibit an explicit (non--K\"ahler and non--Einstein)
metric on the latter. Section 8 concludes with a comparison of the 
geometrical results obtained here with the results of $a$--maximisation 
applied to the quiver gauge theory corresponding to the complex cone 
over the first del Pezzo surface \cite{BB}. 
In Appendix A the techniques used in the paper, 
which perhaps are unfamiliar to many physicists, are applied to the
familiar example of the conifold.


\section{Sasaki--Einstein Metrics on $S^2 \times S^3$}
\label{section2}

In this section we review the geometry of the recently discovered 
Sasaki--Einstein metrics on $S^2 \times S^3$ \cite{paper2}. There
is an infinite family of such metrics, labeled by two coprime
integers $p>1$, $q<p$ -- we refer to these as $Y^{p,q}$. Geometrically
they are all $U(1)$ principle bundles\footnote{This $U(1)$ is {\bf not}
to be confused with the isometry generated by the Reeb vector. The latter
is embedded non--trivially inside the torus defined by this $U(1)$ and
and $U(1)$ that rotates the axially squashed $S^2$ fibre.}
over an axially squashed $S^2$ bundle over a round $S^2$. The integers
label the twisting, or Chern numbers, of the $U(1)$ bundle over the
two two--cycles, with the constraint $q<p$ arising as a regularity
condition on the metric. The manifolds are
all cohomogeneity one. The fact that they are all topologically
$S^2 \times S^3$ follows from a theorem of Smale \cite{smale}
on the classification of five--manifold topology.

In the following we first recall basic material about Sasakian--Einstein geometry
and then turn to the metrics $Y^{p,q}$.

\subsection{Sasakian--Einstein geometry}

A Sasaki--Einstein manifold may be defined as a complete positive
curvature Einstein manifold\footnote{We also require
  simply--connectedness. This is not strictly necessary. However,
  given this condition we can
  use a theorem which relates contact structures to
  the existence of globally--defined Killing spinors. The latter is the
physical property that we wish our manifolds to possess.}
 whose metric cone is Ricci--flat K\"ahler
{\it i.e.} a Calabi--Yau cone. The structure of a Sasaki--Einstein
manifold may thus be thought of as ``descending'' from the Calabi--Yau
structure of its metric cone (\ref{metcone}). In particular, 
contracting the Euler vector
$r\partial/\partial r$, which generates the homothetic $\R^+$ action on
the cone, into
the K\"ahler form gives rise to a one--form on the base of the cone, $Y$.
The dual of this is a constant norm Killing vector field -- called
the  Reeb vector in the mathematical literature -- which via
the AdS/CFT correspondence is isomorphic to the R--symmetry of the
dual field theory. The Killing vector defines a foliation of the
Sasaki--Einstein manifold, and one finds that the
transverse leaves have a K\"ahler--Einstein structure. More precisely, 
one can write the \emph{local} form of the metric as follows:
\bea
\label{SE}
   \diff s^2(Y) & = & \diff s^2_4+ \left(\tfrac{1}{3}\diff\psi' + \sigma\right)^2
\eea
where $\diff s^2_4$ is a local K\"ahler--Einstein metric. In particular we
have that
\bea
\label{KEstructure}
\diff \sigma & = & 2 J_4\nonumber\\
\diff \Omega_4 & = & i 3 \sigma \wedge \Omega_4
\eea
where $J_4$ and $\Omega_4$ are the local K\"ahler and holomorphic $(2,0)$
form for $\diff s^2_4$, respectively. The Reeb Killing vector is given by
\bea
\label{sasvec} K \equiv 3\frac{\partial}{\partial \psi'}~.
\eea

Sasaki--Einstein manifolds may then be classified into three
families, according to the global properties of the orbits of this
Killing vector field:
\begin{itemize}

\item If the orbits close, and moreover the associated $U(1)$ action
is free, the Sasaki--Einstein manifold is said to be {\bf regular}. The
length of the orbits are then all equal. One thus has a principle $U(1)$ 
bundle over a four--dimensional base K\"ahler--Einstein manifold.

\item Suppose that the isotropy group $\Gamma_x$ of at least one point $x$ is
non--trivial. Notice that $\Gamma_x$ is necessarily isomorphic
to $\Z_m$, for some integer $m$, since these are
 precisely the proper subgroups of $U(1)$. The $U(1)$ action is then
locally free, meaning that the isotropy groups are all finite --
note that the Killing vector cannot vanish anywhere since it has constant
norm. The Sasaki--Einstein
manifold is then said to be
{\bf quasi--regular}. In this case notice that the length of the orbit
 through $x$ is $1/m$ times the length of the generic orbit.
The quotient of any manifold by a locally free compact Lie group action
is canonically an orbifold. One thus has
 a principle orbifold $U(1)$
bundle, or orbibundle, over a K\"ahler--Einstein base
orbifold. Moreover, the point $x$ will
descend to a $\Z_m$--orbifold point $x$ in this base space.

\item If the orbits do not close, the Sasaki--Einstein manifold is
said to be {\bf irregular}. In this case one does not have a well--defined 
quotient space. Note that such a Sasaki--Einstein manifold
necessarily has at least a $U(1)^d$ isometry group, $d\geq 2$, with
the orbits of the Killing vector filling out a dense subset of the
orbits of the torus action. Indeed, the isometry group of a compact 
Riemannian manifold is always a compact Lie
group.  Hence the orbits of a Killing vector field define a one--parameter
subgroup, the closure of which will always be an abelian subgroup 
and thus a torus. The dimension of the closure of the orbits
is called the \emph{rank}. Thus irregular Sasaki--Einstein manifolds have
rank greater than $1$.

\end{itemize}

\noindent
The five--dimensional regular Sasaki--Einstein manifolds are
classified completely~\cite{fried}. This follows since the smooth
four--dimensional K\"ahler--Einstein metrics with positive
curvature on the base have been classified by Tian and Yau
\cite{tian,tianyau}. These include the special cases $\cp^2$ and
$S^2\times S^2$, with corresponding Sasaki--Einstein manifolds
being the homogeneous manifolds $S^5$ (or $S^5/\Z_3$) and
$T^{1,1}$ (or $T^{1,1}/\Z_2$), respectively. 
For the remaining metrics, the base is a del Pezzo surface
obtained by blowing up $\cp^2$ at $k$ generic points with $3\le
k\le8$ and, although proven to exist, the generic metrics are not 
known explicitly.

We emphasise the lack of existence of K\"ahler--Einstein metrics on the del
Pezzo surfaces with one or two points blown up, as this will play an important role
later. This fact is actually rather simple to understand.
It is a fairly straightforward calculation \cite{matsushima} 
to show that the Lie algebra
$\mathtt{H}$ generated by
holomorphic vector fields on a K\"ahler--Einstein manifold is a
complexification of the Lie algebra generated by Killing vector fields
{\it i.e.} isometries. The latter is always a reductive algebra (meaning
it is the sum of its centre together with a semi--simple algebra)  but
for the first and second del Pezzo surfaces the algebra $\mathtt{H}$ is
not reductive. Clearly then $\mathtt{H}$ being reductive is
always necessary. This is Matsushima's Theorem \cite{matsushima}. 
One also requires that the anti--canonical bundle 
be ample, that is $c_1>0$, otherwise the putative K\"ahler--Einstein metric 
would be indefinite. In complex dimension two, these necessary 
conditions are in fact
sufficient for existence of a K\"ahler--Einstein metric \cite{tian,tianyau},
and this leads to the list stated above.

It was only recently
~\cite{boyerone,Boyer:2000pg,boyerthree,boyerfour} that
quasi--regular Sasaki--Einstein metrics were shown to exist on
$\#l(S^2\times S^3)$ with $l=1,\dots, 9$. In particular, there are
$14$ known inhomogeneous Sasaki--Einstein metrics on $S^2\times S^3$. We stress that
the proof of this is via existence arguments, rather than giving
explicit metrics. Specifically, one uses a modification of Yau's
argument to prove existence of K\"ahler--Einstein metrics on
certain complex orbifolds, and then builds the appropriate
$U(1)$ orbibundle over these to obtain Sasaki--Einstein manifolds. One
can also obtain quasi--regular geometries rather trivially by taking
quotients of the explicit regular geometries discussed above
by appropriate freely--acting finite groups. For example, one can
take a freely--acting finite
subgroup of $SU(3)$ and quotient $S^5\subset \C^3$ by the induced action.

\subsection{The metrics $Y^{p,q}$}

We will now review, as well as work out some new, properties of the 
Sasaki--Einstein metrics $Y^{p,q}$ on $S^2\times S^3$. These  were presented  in
\cite{paper2} in the following local form:
\bea
\label{tinky}
  \diff s^2 &=& \frac{1-cy}{6}(\diff\theta^2+\sin^2\theta
      \diff\phi^2)+\frac{1}{w(y)q(y)}
      \diff y^2+\frac{q(y)}{9}(\diff\psi-\cos\theta \diff\phi)^2 \nonumber\\ 
      & + &  {w(y)}\left[\diff\alpha +f(y) (\diff\psi-\cos\theta
      \diff\phi)\right]^2\nn\\ 
      & \equiv & \diff s^2(B)+w(y)[\diff\alpha+A]^2
\eea
where
\bea
w(y) & = & \frac{2(a-y^2)}{1-cy} \nn\\ 
q(y) & = & \frac{a-3y^2+2cy^3}{a-y^2} \nn\\ 
f(y) & = & \frac{ac-2y+y^2c}{6(a-y^2)}~.
\eea
For $c=0$ the metric takes the local form of the standard
homogeneous metric on $T^{1,1}$. Otherwise, $c$ can be scaled to
1 by a diffeomorphism. Henceforth we assume this is the case.

\subsubsection*{The base $B$}

The analysis of \cite{paper2} first showed that the four dimensional space $B$ can be
made
into a smooth complete compact manifold with appropriate choices for the ranges of the coordinates.
In particular, for\footnote{In the limit $a\rightarrow1$ the two
positive roots become equal and $y=1$ is a double root. In the case
$a=1$ the metric is locally that of the round metric on $S^5$. }
\bea 
0<a<1~, 
\eea
one can take the ranges of the
coordinates $(\theta,\phi,y,\psi)$ to be $0\le\theta\le \pi$,
$0\le \phi \le 2\pi$, $y_1\le y\le y_2$, $0\le \psi \le 2\pi$ so
that the ``base space'' $B$ is an axially squashed $S^2$ bundle
over a round $S^2$. The latter is parametrised by $\theta$,
$\phi$, with $\psi$ being an azimuthal coordinate on the axially
squashed $S^2$ fibre. This bundle is geometrically twisted, and
may be thought of as the $S^2$ bundle over $S^2$ formed by taking the
tangent bundle of the round two--sphere and adding a point at infinity
to each fibre. Now, the
inclusion map $U(1)\hookrightarrow SO(3)$ induces a map $\Z\cong
\pi_1(U(1)) \rightarrow \pi_1(SO(3))\cong \Z_2$ which is reduction
modulo $2$. Here we are thinking of $U(1)$ as the group in which the
transition functions of $TS^2$ take their values, and $SO(3)$ as the
structure group of the associated oriented $S^2$ bundle over $S^2$.
Since $TS^2$ has Chern number $2 \cong 0 $ mod $2$, it follows that the $S^2$
bundle is trivial and thus the
manifold $B$ is topologically a product space, $B\cong S^2 \times S^2$.
The range
of $y$ is fixed so that $1-y>0$, $a-y^2>0$, $w(y)>0$, $q(y) \ge
0$. Specifically, $y_i$  are two zeroes of $q(y)$, {\it i.e.}  are
two roots of the cubic
\bea
{\cal Q}(y) \equiv a-3y^2+2 y^3  =  0~.
\label{thecubic}
\eea
If $0<a<1$ there are three real roots, one negative ($y_1$) and two positive, the
smallest being $y_2$. The values $y=y_1,y_2$ then correspond to the south and
north poles of the axially squashed $S^2$ fibre. One may check explicitly
that the metric is smooth here with the above identifications of coordinates.

\subsubsection*{The circle fibration}

It was shown in \cite{paper2} that
for a countably infinite number of values of $a$, with $0<a<1$, one can now
choose the period of $\alpha$ so as to describe a principle $S^1$
bundle over $B$. This is true if and only if the periods
of $\diff A$ are \emph{rationally related}. Thus one requires
\be
P_1 = \ell p, \quad P_2 = \ell q\ee
with the periods $P_i$, $i=1,2$, given by
\bea
   P_i &=& \frac{1}{2\pi}\int_{C_i} \diff A
\eea 
where $C_1$ and $C_2$ give the standard basis for the homology
group of two--cycles on $B\cong S^2 \times S^2$. In this case,
one may take
\bea 0\le\alpha\le2\pi\per \eea
and the five--dimensional space is then the total
space of an $S^1$ fibration over
$B\cong S^2 \times S^2$, with Chern numbers $p$ and $q$ over the two
two--cycles. An explicit calculation shows that
\begin{equation}
\label{ratcond}
   \frac{P_1}{P_2} =~\frac{3}{2(y_2-y_1)}.
\end{equation}
Moreover, the function $y_2(a)-y_1(a)$ is a monotonic increasing
function of $a$, taking the range $0<y_2(a)-y_1(a)<3/2$
thus implying a countably infinite number of solutions with $0<q/p<1$.
Furthermore, for any $p$ and $q$ coprime, the space
$Y^{p,q}$ is topologically $S^2\times S^3$ -- see \cite{paper2}.
This follows from a result of Smale on the classification of
five--manifold topology.

\subsubsection*{The volumes}

One finds that
\begin{equation}
   \label{per-pq}
   \per = \frac{q}{3q^2-2p^2+p(4p^2-3q^2)^{1/2}}
\end{equation}
and the volume of $Y^{p,q}$ is given by
\be
\label{volume}
   \mathrm{vol} (Y^{p,q}) =
     \frac{q^2[2p+(4p^2-3q^2)^{1/2}]}{3p^2[3q^2-2p^2+p(4p^2-3q^2)^{1/2}]}
     \pi^3
\ee
which is a quadratic irrational number times 
the volume $\pi^3$ of a unit round $S^5$. We note that at fixed $p$ the volume
is a monotonic function of $q$, and is bounded by the following values 
\bea
\mathrm{vol} (T^{1,1}/\Z_p) > \mathrm{vol} (Y^{p,q}) >\mathrm{vol} (S^5/\Z_{2}\times \Z_p) ~.
\label{limitvolumes}
\eea
The rational case, which is easily seen to correspond to quasi--regular
 manifolds, is described by $p,q\in
\mathbb{N}$, $\mathrm{hcf}(p,q)=1$, $q<p$, which are solutions to the
quadratic diophantine
\bea
4p^2 - 3q^2 & = & n^2
\label{dio}
\eea
for some $n\in \mathbb{Z}$. The solutions to this were given in
closed form in \cite{paper2}.

\subsubsection*{The isometry group}

The isometry group of the metrics (\ref{tinky}) is clearly
locally $SU(2)\times U(1) \times U(1)$, and in particular there are three
commuting Killing vectors
$\de / \de \phi$, $\de / \de \psi$, and $\de / \de \gamma$. Here
we have defined
\bea
\alpha  & \equiv & \ell \gamma
\eea
so that the three generators have canonical period $2\pi$. For us it will be important to note
that the global form of the \emph{effectively} acting isometry group depends
on $p$ and $q$. In particular, for both $p$ and $q$ odd it is
$SO(3)\times U(1)^2$ otherwise it is $U(2)\times U(1)$. This will be explained
later in Section 4. Note that this is precisely analogous to the case of
the Einstein manifolds known in the physics literature as $T^{p,q}$. For
these the effectively acting isometry group is shown \cite{WZ} to be $SO(3)\times SU(2)$ when
one integer is even, and $SO(4)\cong (SU(2)\times SU(2)) / \Z_2$
when both are odd. The latter of course includes 
the case of $T^{1,1}$ \cite{KW}.

\subsubsection*{The local K\"ahler--Einstein structure}

Employing the change of coordinates $\alpha=-\beta/6-\psi'/6$,
$\psi=\psi'$ one can \cite{paper2} bring the metric (\ref{tinky}) into the local
Sasaki--Einstein form (\ref{SE}). In particular
\begin{equation}
\label{canmetric}
\begin{aligned}
   \diff s^2 &=  \frac{1-y}{6}(\diff \theta^2+\sin^2\theta \diff \phi^2)
      + \frac{\diff y^2}{w(y)q(y)}
      + \frac{1}{36} w(y)q(y)(\diff \beta+\cos\theta \diff \phi)^2 \\
      & + \frac{1}{9}[ \diff \psi'-\cos\theta \diff \phi
         +y(\diff \beta+\cos\theta \diff \phi)]^2~.
\end{aligned}
\end{equation}
The corresponding $J_4$ and $\Omega_4$, satisfying (\ref{KEstructure}), can be taken as
\bea
J_4 & = & \frac{1-y}{6}\sin\theta \diff \theta \wedge \diff \phi + \frac{1}{6} \diff y
\wedge (\diff \beta +\cos\theta \diff \phi)\label{canj4}\\
\Omega_4 & = & \sqrt{\frac{1-y}{6w(y)q(y)}} (\diff \theta +i \sin\theta \diff
\phi)\wedge \left[\diff y + i \frac{w(y)q(y)}{6}(\diff \beta +\cos\theta
\diff \phi)\right]\label{canomega4}
\eea
while the Reeb Killing vector is given by
\bea
\label{sasvec2} K
& = & 3\frac{\partial}{\partial \psi}-\frac{1}{2\ell}
\frac{\partial}{\partial \gamma}~.
\eea
Note that this has compact orbits when $\per$ is a rational number
and corresponds to the quasi--regular class, by definition.
This is true if and only if (\ref{dio}) holds.
If $\per$ is irrational the generic 
orbits do not close, but instead densely fill the
orbits of the torus generated by $[\partial/\partial\psi$,
$\partial/\partial\gamma ]$ and we thus fall into the
irregular class. The rank of these metrics is thus equal to 2.
Note that the orbits close only over the 
submanifolds given by $y=y_1,y_2$. These are precisely the 
special\footnote{The manifolds $Y^{p,q}$ are cohomogeneity one, meaning 
that the generic orbit under the action of the isometry group is 
codimension one. There are then always precisely two special orbits of 
higher codimension.} orbits of the cohomogeneity one action.

\subsubsection*{The Killing spinors}

To show that these manifolds admit globally defined Killing spinors one
appeals to the following theorem \cite{boyer3sas}: every simply--connected
spin Sasaki--Einstein manifold, where the latter is defined in terms
of the existence of
a certain contact structure, admits a solution to the Killing spinor
equation. In particular we note that the dual one--form to $K$ is given
by
\bea
\eta & =&  -2y(\diff\alpha + A) + \frac{1}{3}q(y)(\diff\psi-\cos\theta\diff\phi)
\eea
which is globally--defined (the factor of  $q(y)$ is essential here).
The contact structure is then easy to
exhibit in terms of $\eta$ for the manifolds
$Y^{p,q}$ \cite{paper2}. This theorem is the reason
why one {\it a priori} requires $\mathrm{hcf}(p,q)=1$ -- however see below.

\subsubsection*{The Calabi--Yau cones}

It will be important for us to exploit the symplectic
structure of the associated Calabi--Yau cones. Rather generally, the
Calabi--Yau structure on the metric cone is
specified by a K\"ahler (hence also symplectic)
form $J$ and a holomorphic $(3,0)$ form $\Omega$, which in terms of
the four--dimensional K\"ahler--Einstein data read as follows:
\bea
J
& = &  r^2 J_4 +r \diff r
\wedge (\tfrac{1}{3}\diff\psi'+\sigma)\label{CYJ}\\
\Omega & = & e^{i\psi'} r^2 \Omega_4 \wedge \left[ \diff r + ir
(\tfrac{1}{3}\diff \psi' +\sigma)\right]~.
\label{omegacycone}
\eea
In the specific case of $C(Y^{p,q})$, we have
\bea 
J & = & r^2 \frac{1-y}{6}\sin\theta\diff\theta\wedge\diff\phi \nn \\   
&+ & \frac{1}{3}r\diff r\wedge (\diff\psi-\cos\theta\diff\phi) 
- \diff (yr^2)\wedge \left(\diff\alpha
+\frac{1}{6}(\diff\psi-\cos\theta\diff\phi)\right)\label{sympYpq}\eea
and
\bea
\Omega & = & e^{i\psi} r^2 \sqrt{\frac{1-y}{6w(y)q(y)}} (\diff\theta +
i \sin\theta\diff\phi )\nonumber\\
&\wedge  & 
\left[\diff y -iw(y)q(y)\left(\diff \alpha
+\tfrac{1}{6}(\diff\psi-\cos\theta\diff\phi)\right)\right]\nn\\ 
& \wedge & \left[\diff r - 2ir \left(y \diff \alpha +(y-1)\tfrac{1}{6}(\diff\psi 
-\cos\theta\diff\phi)\right)\right]
\eea
where we used (\ref{canj4}) (\ref{canomega4}) 
and have then rewritten the expressions in terms of the original coordinates.

Note that this calculation  
shows that $\Omega$ is invariant under $\de/\de\alpha$,
namely
\bea
{\cal L}_{\de/\de\alpha} \Omega = i_{\de/\de \alpha}\diff \Omega + 
\diff (i_{\de/\de \alpha} \Omega ) = 0 
\eea
implying that the Killing spinors are also invariant. This explicitly checks that 
upon performing a $T$--duality along the $\alpha$ direction to Type IIA string theory,
the number of preserved supersymmetries is unchanged. 
In fact, this is 
obvious given the original construction \cite{paper1} of these metrics. 
Since we are guaranteed existence of Killing spinors by the 
theorem of \cite{boyer3sas}, and since we have now shown that the 
spinors are independent of $\alpha$, it follows that one may in fact take 
$\mathrm{hcf}(p,q)=h>1$ by taking a smooth quotient by $\Z_h$ of 
the simply--connected Sasaki--Einstein manifold 
$Y^{p/h,q/h}$. Since this is rather trivial, we 
take this as understood in the remainder of the paper.

\subsubsection*{Complex coordinates}

It is easy to introduce a (local) set of complex coordinates. To do so 
we seek three closed complex one--forms $\eta^i$ such that $\Omega \wedge \eta^i=0.$ First,
consider the following local one--forms obeying the latter property:
\bea
\label{fuckbush}
\eta^1 & = & \frac {1}{\sin\theta} \diff \theta + i \diff \phi \nn \\ 
\tilde{\eta}^2 & = & \frac {1}{w(y)q(y)}\diff y - i (\diff \alpha + \frac
{1}{6}(\diff\psi-\cos\theta\diff\phi))\nn\\ 
\tilde{\eta}^3 & = & \frac {\diff r }{2r} - i \left( \diff \alpha + (y-1)(\diff \alpha + \frac
{1}{6}(\diff\psi-\cos\theta\diff\phi))\right)
\eea
where now 
\bea
\Omega & = & 2 e^{i\psi}r^3\sqrt{\frac{{\cal Q}(y)}{3}}\sin\theta
\eta^1 \wedge \tilde{\eta}^2 \wedge \tilde{\eta}^3~.\label{omegaetas}
\eea
Taking $z_1  =  \tan\tfrac{\theta}{2}e^{i\phi}$ we immediately find
\bea
\eta^1 & = &  \frac{\diff z_1}{z_1}~.
\eea
To obtain two more integrable one--forms one is free to
consider linear combinations of the one--forms (\ref{fuckbush}). Take
\bea
\eta^2 & = & -\frac{1}{6}\cos\theta\eta^1 + \tilde{\eta}^2\nn \\
\eta^3 & = &  \frac{1}{6} \cos\theta\eta^1-y \tilde{\eta}^2 + \tilde{\eta}^3~.
\eea
Notice that one can now simply drop the tildes in 
(\ref{omegaetas}). Moreover the $\eta^2$, $\eta^3$ are now closed and hence locally 
exact. In particular 
\bea
\eta^i & = & \frac{\diff z_i}{6z_i}
\eea 
$i=2,3$,  with
\bea
z_2 & = & \frac{1}{\sin\theta}
\sqrt{ (y-y_1)^{-\frac{1}{y_1}} (y_2-y)^{-\frac{1}{y_2}} (y_3-y)^{-\frac{1}{y_3}}}
~e^{-6i\alpha-i\psi} \nn \\ 
z_3 & = & r^3 \sin\theta \sqrt{{\cal Q}(y)}~e^{i\psi}~.
\eea
In terms of the $\{z_i\}$, the three--form assumes a very simple form:
\bea
\Omega & = & \frac{1}{18\sqrt 3} \frac{\diff z_1
\wedge \diff z_2 \wedge \diff z_3}{z_1 z_2}~.
\eea

\subsubsection*{Supersymmetric cycles}

In this subsection we will show that the cones over the submanifolds 
$y=y_1,y_2$, which recall are the special orbits of the cohomogeneity 
one action, are in fact \emph{divisors} in the Calabi--Yau cone. 
This amounts to showing that they are calibrated with respect to the four-form
$\tfrac{1}{2}J\wedge J$. We denote the three--submanifolds as $\Sigma_i$, 
$i=1,2$, respectively.

Thus, we compute the pull--back of $\tfrac{1}{2}J\wedge J$ to the four--cycles
in the Calabi--Yau cone $C(Y^{p,q})$ specified by $y=y_i$. 
The latter are in fact cones over the Lens spaces $\Sigma_1\cong S^3/\Z_{p+q}$, $\Sigma_2\cong S^3/\Z_{p-q}$. We shall 
show in detail that this is indeed the topology in Section \ref{momentmapsec}. However,
this fact can also be seen by computing the pull--back of the K\"ahler form 
to the four--submanifolds. Defining $k=p+q$, $l=p-q$, these 
are\footnote{Recall that $y_1<0$ and $y_2>0$.}
\bea
J|_{y=y_1}&  = & \ell y_1 \left[ -\frac{k}{2}r^2 \sin\theta \diff \theta \wedge \diff \phi
- r \diff r \wedge (\diff 2\gamma - k \cos\theta \diff \phi)\right]\\
 J|_{y=y_2}&  = & \ell y_2 \left[ \frac{l}{2}r^2 \sin\theta \diff \theta \wedge \diff \phi
- r \diff r \wedge (\diff 2\gamma +l  \cos\theta \diff \phi)\right]
\eea
and are precisely the K\"ahler forms associated to cones over round Lens spaces $S^3/\Z_k$
and $S^3/\Z_l$, respectively. Indeed, since $\gamma$ has period $2\pi$, 
the one--forms multiplying $\diff r$ are precisely global angular 
forms (global connections) on the total spaces of circle bundles over 
$S^2$ with Chern numbers $k$ and $-l$, respectively. The total spaces of 
such bundles are precisely $S^3/\Z_k$ and $S^3/\Z_l$, respectively.
 From these expressions, one calculates
\bea
\frac{1}{2} J \wedge J|_{y=y_i} & = & \frac{r^3\ell y_i(1-y_i)}{3}\sin\theta\diff\theta 
\wedge\diff\phi\wedge \diff\gamma\wedge \diff r~.\label{Jsquare}
\eea
Let us compare this with the volume form induced on $\Sigma_i$ from
the metric (\ref{tinky}). This is given by
\bea
\mathrm{vol} & = & 
\frac{r^3 \ell \sqrt{w(y_i)}(1-y_i)}{6}\sin\theta\diff\theta\wedge\diff\phi \wedge
\diff\gamma \wedge \diff r~.
\eea
Remarkably, since $w(y_i)=4y_i^2$ at any root of the cubic (\ref{thecubic})
 we see that this precisely agrees
with (\ref{Jsquare}). Thus we see that both $C(\Sigma_1)=\{y=y_1\}$ and 
$C(\Sigma_2)=\{y=y_2\}$ are 
divisors of $C(Y^{p,q})$, or in other words they are
supersymmetric submanifolds.

We may now write down the volumes of the $\Sigma_i$. Here one needs to 
use the explicit formulae for the roots of the cubic 
$y_1$ and $y_2$ in terms of $p$ and $q$:
\bea
y_1 & = & \frac{1}{4p}\left(2p-3q-\sqrt{4p^2-3q^2}\right)\nn\\  
y_2 & = & \frac{1}{4p}\left(2p+3q-\sqrt{4p^2-3q^2}\right)~.
\eea
One then easily calculates
\bea\label{vols}
\mathrm{vol}(\Sigma_1) & = &
\frac{q^2(p+q)\left(-2p+3q+\sqrt{4p^2-3q^2}\right)^2}{2p^2\left(3q^2-2p^2+p\sqrt{4p^2-3q^2}\right)^2}\pi^2\nn \\ 
\mathrm{vol}(\Sigma_2) & = &
\frac{q^2(p-q)\left(2p+3q-\sqrt{4p^2-3q^2}\right)^2}{2p^2\left(3q^2-2p^2+p\sqrt{4p^2-3q^2}\right)^2}\pi^2
~.\eea
In particular, let us write down the volumes of $\Sigma_i$ in the 
case of $p=2$, $q=1$:
\bea
\mathrm{vol} (\Sigma_1) =  \frac{\pi^2}{108} (31+7\sqrt{13})~,\qquad \qquad
\mathrm{vol} (\Sigma_2) =  \frac{\pi^2}{36} (7+\sqrt{13})~. \label{y21vols}
\eea

\section{Moment maps and convex rational polyhedral cones}
\label{section3}

In the remainder of this paper it will be crucial for us that the
Sasaki--Einstein manifolds $Y^{p,q}$ admit an effectively acting
three--torus $\T^3=U(1)^3$ of isometries, which moreover is
\emph{Hamiltonian}. The latter means that the
action preserves the symplectic form of the cone $C(Y^{p,q})$
and that one can use this to introduce a moment map. The
torus is just the maximal torus in the isometry group,
and the fact that the torus is half the dimension of the
cone means that, by definition, the cones are \emph{toric}. The image
of the cone under the corresponding moment map generally belongs to
a special class of convex rational
polyhedral cones in $\R^3$ \cite{FT, L} -- these are simply convex cones formed
by intersecting some number of planes through the origin. The
normal vectors to these planes, or \emph{facets}, are necessarily
rational and describe which $U(1)$ subgroup of $\T^3$ is vanishing
over the corresponding codimension two submanifold of
$C(Y^{p,q})$. This generalises the well--known result in symplectic
geometry that the image of the moment map for a compact toric
symplectic manifold is always a particular type of convex rational
polytope called a \emph{Delzant} polytope. 

In this section we give a general review of symplectic toric geometry. 
This is mainly rather standard material from the point of view 
of a symplectic geometer -- the reader who is familiar with this
subject may therefore wish to skip this section. On the other hand, 
we hope that this will be a useful
self--contained presentation of the material.

\subsection{Moment maps for torus actions}

In this subsection we give a general summary of moment maps,
Hamiltonian torus actions, and symplectic toric manifolds, orbifolds
and cones,
together with the properties of their images under the moment maps, which
are always particular types of rational polytopes (or polyhedral cones)
in $\R^n$.
The case of compact manifolds \cite{atiyah,GS,D}
is rather standard in symplectic geometry,
but the generalisation for orbifolds \cite{LT}, and especially cones
\cite{FT, L}, is quite recent. 

We begin by giving a general definition. Suppose that the torus
$\T^n$ acts effectively -- meaning that every non--trivial
element moves at least one point -- on a symplectic manifold $M$
with symplectic form $\omega$. We identify the Lie algebra of this
torus, as well as its dual, with Euclidean $n$--space, so
$\mathsf{t}_n\equiv \mathsf{Lie}(\T^n)\cong\R^n$,
$\mathsf{t}_n^*\cong \R^n$. Then a moment map for the torus action
is simply a $\T^n$--invariant map
\be \mu:M\rightarrow
\mathsf{t}_n^*\cong \R^n\ee
satisfying the condition
\be
\diff\mu^{i} = V^i\lrcorner\omega~.\ee
Here $V^i$ denotes the vector field on $M$
corresponding to the basis vector $e_i$ in $\mathsf{t}_n\cong \R^n$, and $\mu^{i}$ denotes the component of the map $\mu$ in the
direction $e_i$ {\it i.e.} $\mu = (\mu^{1},\ldots,\mu^{n})$. Clearly this moment map is unique only up to an
additive integration constant. 

To see where this map comes from,
suppose for simplicity that one has a $U(1)$ action on a symplectic manifold $M$, generated by some vector field $V$, which
moreover preserves the symplectic form. One then says that the $U(1)$ action
is \emph{symplectic}. The latter means that
\be
\mathcal{L}_{V}\omega=0\ee
where $\mathcal{L}$ is the Lie derivative. Since $\omega$ is closed, this condition is just
\be
\diff(V\lrcorner\omega)=0~.\ee
As long as the closed one--form $V\lrcorner\omega$ is trivial as a
cohomology class, $[V\lrcorner\omega]=0\in H^1(M;\R)$,
then one can ``integrate'' this equation to a function $\mu$,
which is precisely the moment map for the $U(1)$ action. The
action is then said to be \emph{Hamiltonian}. For example,
the $U(1)$ which rotates one of the circles in $\T^2$, with
obvious symplectic form, is \emph{not} Hamiltonian. Clearly, if
$H^1(M;\R)$ is trivial then all symplectic actions are in fact Hamiltonian.
A \emph{symplectic toric manifold} is then by definition a symplectic manifold
of dimension $2n$ with an effective Hamiltonian torus action by $\T^n$.

It is by now a classic fact in symplectic geometry that, for a compact symplectic toric manifold $M$, the
image of $M$ under $\mu$ is a certain kind of convex rational polytope in $\R^n$ called a \emph{Delzant
polytope} \cite{D}. Recall that a polytope is just the convex hull of some finite number of points in $\R^n$. The
codimension one hyperplanes that bound the polytope are called its facets.
The symplectic toric manifold is then a torus fibration over this
polytope, with the fibres collapsing in a certain way over the facets.
More precisely, over an interior
point of the polytope the fibre of the moment map (the inverse image of
the point)
is the whole torus $\T^n$, but over the boundary facets
this fibre collapses to $\T^{n-1}\cong\T^n/U(1)$. Such a $U(1)$ subgroup
is specified by a vector in the weight
lattice $v\in\Z^n$ of $\T^n$, and this vector is in fact just the normal vector to the
facet. Moreover the $U(1)$ fixes a corresponding
codimension two submanifold of $M$. To see this, consider
the case
where $v=e_1 = (1,0,\ldots,0)$. Denote the corresponding vector field as $V$. Then
over a codimension two fixed point set $F\subset M$ we have that $V=0$, and moreover $F$ is
itself symplectic toric with respect to the torus $\T^{n-1}\cong \T^n/U(1)$. In particular, the moment map
$\mu$ restricted to $F$ is constant in the direction corresponding to $V$        {\it i.e.}  $\mu^1=c=\mathrm{constant}$.
Then $\mu_F\equiv \mu\mid_{F}$ is a moment map for $F$
with $<\mu_F, e_1>=c$. This defines the hyperplane
at $x_1=c$, where $\{x_i\}$, $i=1,\ldots,n$ are coordinates on $\R^n$. The general case follows similarly.
The normal vectors to the facets  are thus all
\emph{rational} vectors. If two facets intersect over a codimension two face in $\R^n$, then
both the corresponding $U(1)$'s vanish, and the fibre over this face is a $\T^{n-2}$. Continuing in this way,
the vertices of the polytope are precisely the points in $M$ which are fixed under the entire torus
action. The fact that the polytope is always convex follows from an
argument using Morse theory \cite{atiyah, GS}.

Delzant polytopes satisfy some additional conditions, as well as being rational:%
\begin{itemize}
\item \emph{simplicity} -- $n$ edges meet at each vertex.
\item \emph{smoothness} -- for each vertex, the corresponding $n$
edge vectors $u_i$, $i=1,\ldots,n$ form a $\Z$--basis\footnote{This
means that the set $\{\sum_i n_iu_i\mid n_i\in \Z, i=1,\ldots,n\}$ is precisely
$\Z^n$.}
of $\Z^n$.
\end{itemize}
The polytope data is sufficient to recover the original
symplectic toric manifold. Moreover, the correspondence between Delzant
polytopes and compact symplectic toric manifolds is one--to--one. Thus,
to any Delzant polytope $\Delta$ one can associate a corresponding
symplectic toric manifold whose image under the moment map is precisely $\Delta$.
The proof of this is by construction. This will be
extremely important for us in Section 5 -- in physics terms, the
construction realises the manifold as the vacuum of a gauged linear
sigma model \cite{wittenLSM}.

We now briefly explain the above conditions. Assuming the first
condition holds, the second condition avoids orbifold
singularities. Indeed if the smoothness condition fails then
$\T^n/<u_i>\cong \Gamma$ is a non--trivial finite abelian group, where
$<u_i>$ denotes the span of the $u_i$ over $\Z$. In this case the
corresponding point in $M$ is an orbifold point with structure
group $\Gamma$. Indeed, there is a corresponding classification of
symplectic toric orbifolds where the smoothness condition
is dropped, and moreover one attaches to each facet a positive
integer label \cite{LT}. This latter necessity can be seen by
considering the weighted projective space $\cp^1_{[k,l]}$. This is
topologically a sphere, with neighbourhoods of the north and south
poles replaced by orbifold singularities $\C/\Z_k$ and $\C/\Z_l$,
respectively. The quotient by the $U(1)$ action which rotates
around the equator is clearly just a line segment. Thus the
orbifold information is completely lost when one takes the image
under the moment map. To remedy this \cite{LT}, quite generally, one
associates to each facet a positive integer label $m$, such that
the pre--image of any point in that facet has local orbifold
structure group $\Z_m$. In the case at hand, the endpoints of the
interval are assigned labels $k$ and $l$, respectively.

The first condition -- simplicity -- avoids even worse singularities than orbifold singularities. As we
shall see, for symplectic toric cones this condition is \emph{not} satisfied
at the vertex corresponding to the apex of the
cone, unless of course the cone is in fact an orbifold singularity.

\subsection{Toric Calabi--Yau cones}

This brings us to the generalisation of this theorem \cite{FT, L}
for \emph{symplectic toric cones}, which is the case of
interest for us. These may be regarded as non--compact symplectic
toric manifolds with a homothetic action of $\R^+$ which commutes with the
torus action and acts by rescaling the symplectic form.
In fact, every symplectic toric cone is a cone
over a toric contact manifold $Y$, and vice versa. In this case the moment map for the symplectic toric cone
$C(Y)=\R^+\times Y$
may still be defined, away from the apex of the cone, and takes a special form. Define
the one--form
\be
\eta_C=r\partial/\partial r \lrcorner \omega\ee
where $r\partial/\partial r$ is the \emph{Euler vector}, which generates
the $\R^+$ action on the cone, and $\omega$ is the symplectic form.
Identifying the base of the cone $Y = C(Y)\mid_{r=1}$ we may define
the one-form\footnote{More precisely we embed $Y$ in $C(Y)$ at $r=1$
and then pull back $\eta_C$ to $Y$ to give $\eta$.}
$\eta=\eta_C\mid_{r=1}$. One then easily sees that
\be
\omega = r\diff r \wedge \eta + \frac{1}{2}r^2\diff\eta~.\ee
A straightforward calculation then shows that the moment map $\mu$ on the cone
is given by
\be
<\mu, e_i> = \eta_C(V^i)\ee
for any basis vector $e_i$ of $\mathsf{t}_n$ and corresponding vector
field $V^i$. Here $\eta_C(V^i)$ just denotes the dual pairing between
one--forms and vectors. The choice of integration constant makes this moment map
transform homogeneously under the $\R^+$ homothetic action. It also ensures that
the apex of the cone, at $r=0$, is mapped to the origin of $\R^n$.

Let us now also assume\footnote{The symplectic toric cones that are \emph{not}
of Reeb type are rather uninteresting: they are either cones over
$S^2\times S^1$, cones over 
principle $\T^3$ bundles over $S^2$, or cones over products $\T^m\times S^{m+2j-1}$,
$m>1, j\geq 0$ \cite{L}.} that the symplectic toric cones are of
\emph{Reeb type}. This means that there is some element $\zeta \in
\mathsf{t}_n\cong\R^n$ such that $<\mu,\zeta>$ is a strictly positive
function on $C(Y)$.
The image of the moment map is then a
\emph{strictly convex rational polyhedral cone} in $\R^n$ \cite{FT}, which, moreover,
is \emph{good} in the sense of reference \cite{L}. Recall that a
rational polyhedral cone
may be defined as a set of points in $\R^n$ of the form
\be
C = \{x \in \R^n\mid <x,v_i>\leq 0, \ i=1,\ldots, d\}\label{polycone}
\ee
where the rational vectors $v_i$ are the outward pointing
normal vectors to
the facets of the cone $C$. Here we may assume
that the set $\{v_i\}$ is
\emph{minimal}, meaning that one cannot drop any vector $v_i$ from the
definition without changing the cone, and also \emph{primitive} -- recall that a
 vector with
integral entries is said to be primitive if it cannot be written
as $n v$ where $1\neq n\in \Z$ and $v$ is also a vector with integral
entries. The requirement that this polyhedral cone
 is strictly convex means that it is
a cone over a polytope.

The ``conelike" nature of the
subspace (\ref{polycone}) of course descends from the ``conelike" nature of the cone we began
with -- the property that
$C(Y)$ is invariant under a group $\R^+$ of homotheties will
be inherited by the image under the moment map since by definition
the moment map commutes
with the $\R^+$ action.
Clearly the
simplicial condition will fail at the apex$=$origin of $\R^n$ unless $d=n$. Moreover,
even in this case the smoothness condition will fail unless
the edges span $\Z^n$. In this case, by an $SL(n;\Z)$ transformation
of the torus, one can take this
to be the standard basis, whence it is easy to see that the cone one
started with is just
$\R^{2n}$ with its usual symplectic structure. This latter point
brings up an issue worth stressing: one is of course free to make
an $SL(n;\Z)$ transformation of the torus $\T^n$ resulting in a change
of the basis $e_i$. This will generate a
corresponding $SL(n;\Z)$ transformation on the image under the moment map. Thus
the polytopes and polyhedral cones are only unique up to such
transformations.

As shown in \cite{L}, the image of a symplectic toric cone under its
moment map is also a \emph{good} polyhedral cone. This means the following. Let
$\mathcal{F}$ be a proper face of the cone $C$. Over this face there will be a corresponding
torus $\T_{\mathcal{F}}\subset \T^n$ which is collapsing to zero. For example, in the case
that $\mathcal{F}$ is a facet, $\T_{\mathcal{F}}\cong U(1)$. For a face
$\mathcal{F}$ of codimension $m$ the
torus is dimension $m$:  $\dim \T_{\mathcal{F}} = m$. Now, the torus
$\T_{\mathcal{F}}\subset \T^n$ determines a
lattice $\Z_{\T_{\mathcal{F}}}= \mathrm{ker}(\exp:\mathsf{t_{\mathcal{F}}}\rightarrow \T_{\mathcal{F}})\subset \Z^n$.
We then require that the corresponding
collection of normal vectors
form an integral basis for this lattice, {\it i.e.} the collection of normal
vectors span the lattice $\Z_{\T_{\mathcal{F}}}$ over $\Z$. This condition may be
regarded as a generalisation of
Delzant's conditions for symplectic toric manifolds to symplectic toric
cones.

In the particular case where the symplectic cone came from a
Calabi--Yau cone, one has additional information. In particular,
the Sasaki--Einstein metric on $Y$ may be used to define the
dual vector field $K$ with $\eta(K)=1$. This is called the
Reeb vector in the language of contact geometry.
Physically this is dual to the R--symmetry of the field theory.  Then
there is a corresponding Lie algebra element
$\zeta\in \mathsf{t}_n$, and we have
\be <\mu_Y, \zeta> = \eta(K) = 1~.\ee
It follows that the image $\mu_Y(Y)$ lies in the above hyperplane,
which is called the \emph{characteristic hyperplane} \cite{boyertoric}.
In particular, note that the polytope one obtains by
intersecting the polyhedral cone with the characteristic hyperplane
will be rational if and only if $\zeta$ is rational.
The latter condition is required precisely for quasi--regularity of
the Sasaki--Einstein metric. Correspondingly, this is also the
condition that the characteristic polytope satisfies
for an orbifold polytope, and thus that the quotient of $Y$ by
the $U(1)$ action generated by $K$ gives an orbifold. Notice that one may then apply the modified
Delzant construction of \cite{LT} to obtain a gauged linear
sigma model describing this orbifold.
In principle one could do this for
our quasi--regular Sasaki--Einstein manifolds, although we will not
pursue this here.

\section{The moment map and its image}
\label{momentmapsec}

In this section we explicitly construct the polyhedral cone corresponding
to the image of $C(Y^{p,q})$ under its moment map.

The Calabi--Yau cones on $Y^{p,q}$ are {\em symplectic toric
cones}. In particular, the $\T^3$ action, which is the maximal
torus of the isometry group, is Hamiltonian, and one can explicitly
integrate the symplectic form (\ref{sympYpq}) to obtain a moment
map. Note in fact that (\ref{sympYpq}) can be written as
\bea
J & = & \diff \phi \wedge \diff \left[ r^2 \frac{1-y}{6}\cos\theta \right]
+ \diff\psi \wedge \diff \left[ -r^2 \frac{1-y}{6} \right] + \diff \gamma \wedge
\diff \left[ \ell r^2 y \right]~.
\eea
The torus $\T^3$ is essentially generated by the Killing vectors
$\de/\de\phi$, $\de/\de\psi$, $\de/\de\gamma$. However, one must be
careful to ensure that the Killing vectors one takes really do form a
basis for an \emph{effectively} acting $\T^3$. Since this is
a slightly subtle point, we first explain a simpler example.

\subsubsection*{A brief detour on Lens spaces}

Let us consider the Lens spaces $L(1,m)=S^3/\Z_m$ where we regard $S^3$ as
a (squashed) Hopf $S^1$ fibration over a round two--sphere.
The isometry groups of the latter may be analysed as follows.
Embed the round sphere $S^3$ in $\R^4$, and regard $\R^4\cong \mathbb{H}$
as the space of quaternions. The isometry group of $S^3$, preserving its
orientation, is $SO(4)\cong (SU(2)_L\times SU(2)_R)/\Z_2$, where $SU(2)_{L,R}$
denote left and right actions by the unit quaternions $Sp(1)\cong SU(2)$. Thus
$\mathbb{H}\ni q \rightarrow aqb^{-1}$ where $(a,b)\in SU(2)\times SU(2)
\cong Spin(4)$. Notice that $(-1,-1)$ acts trivially,
{\it i.e.} the two $SU(2)$ factors intersect precisely over the
antipodal map. Thus, for a
\emph{squashed} three--sphere, meaning that one squashes the Hopf $S^1$
fibre relative to the base round $S^2$, we see that the isometry group
is $U(2)\cong (SU(2)\times U(1))/\Z_2$.

However, suppose we
now take a quotient of $\R^4\cong \mathbb{H}$ on the
right by $\Z_m\subset U(1)$.
One still has a left $SU(2)$ action and a right $U(1)$ action, where the
latter now factors through a cyclic group of order $m$. For example, take
$m=2$, thus
giving $S^3/\Z_2\cong \mathbb{RP}^3$. In complex
coordinates, $\mathbb{H}\cong \C\oplus\C$, this means $(z_1,z_2)\sim
(-z_1,-z_2)$ which identifies antipodal points on the three--sphere. It
follows that the centre of $SU(2)_L$ acts trivially and
hence the effectively acting isometry group is $SO(3)\times U(1)$, where
$U(1)$ rotates the $S^1$ fibre with weight \emph{one} -- half the
weight of $U(1)\subset SU(2)_R$.
It now follows that the isometry group for $S^3/\Z_m$ for all
odd $m$ is $U(2)$, whereas for even $m$ it is $SO(3)\times
U(1)$ -- it is precisely the even cases where
$\Z_m$ contains the antipodal map above.

Clearly these Lens spaces have an isometric $\T^2$ action.
Take $m=2r$. From our discussion above, if $V_1$
denotes the Killing vector that rotates the $S^2$ about its equator
with weight
one, and $V_2$ denotes the Killing vector that rotates the
$S^1$ fibre, also with weight one, then $V_1,V_2$ do indeed
form a basis for an effectively acting $\T^2$. This is the obvious
$\T^2$ in $SO(3)\times U(1)$.

For $m=2r+1$ one needs to be more careful: the isometry group is $U(2)$.
For example, for $r=0$ one has
the unit chiral spin bundle of $S^2$. As is well--known, a single rotation of
$S^2$ will not result in the spinor coming back to itself: one needs to
rotate twice. For an effective action one should thus take a basis
$e_1=V_1+\tfrac{1}{2}V_2$, $e_2=V_2$. Here $e_1$ is half
the generator of the
diagonal $U(1)$ in $SU(2)\times U(1)$, and $V_2$ generates the $U(1)$
factor.

Of course, one can use the basis $e_1=V_1+\tfrac{m}{2}V_2$, $e_2=V_2$ quite generally in
all cases. Indeed, recall that the choice of basis is unique
only up to an $SL(2;\Z)$ transformation. For $m=2r$ even, this basis is just
the $SL(2;\Z)$ transformation
\be
\left(\begin{array}{cc}
1 & r\\
0 & 1\end{array}\right)\ee
of the basis $\{V_1,V_2\}$.

\subsubsection*{The moment cone}

After this brief digression, we return to the case of interest. 
First let us note from the results above that the isometry group of 
the base $B$ is $SO(3)\times U(1)$. Indeed, for fixed $y$, 
$y_1<y<y_2$, we have a copy
of $S^3/\Z_2\cong \mathbb{RP}^3$,
and the group $SO(3)\times U(1)$ acts with cohomogeneity one on $B$ with
fixed $y$ as generic orbit. Thus, in particular, we may take a basis 
$\de/\de\phi+\de/\de\psi$, $\de/\de\phi$ for an effectively acting 
two--torus.
For $C(Y^{p,q})$, one must also add the direction $\de/\de\gamma$. 
However, here one must be careful to ensure the orbits of the vectors 
close, and that this torus then acts effectively, just as for the Lens 
spaces. One finds the following choice suffices:
\bea
e_1 & = & \frac{\de}{\de\phi}+\frac{\de}{\de\psi} \nonumber \\
e_2 & = & \frac{\de}{\de\phi}-\frac{l}{2}\frac{\de}{\de\gamma} \nonumber \\
e_3 & = & \frac{\de}{\de\gamma}~.\label{basis}\eea
Recall that the submanifolds $y=y_1$, $y=y_2$ of $Y^{p,q}$ 
are Lens spaces $S^3/\Z_k$, $S^3/\Z_l$, respectively, where recall
$k=p+q$, $l=p-q$ -- the shift in $e_2$ is then required
precisely by the reasoning above.
Note that one can replace $l$ in the formula for $e_2$ by anything 
congruent to $l$ modulo two (for example, $k$) -- 
this is just an $SL(3;\Z)$ transformation of the torus. Also note that for 
$l$ even one can in fact take a basis $\de/\de\phi$, $\de/\de\psi$, 
$\de/\de\gamma$. The effectively acting 
isometry group is thus $SO(3)\times U(1)\times U(1)$ 
in this case. For $l$ odd this becomes $U(2)\times U(1)$.

Let us now consider the moment map for
$C(Y^{p,q})$. In terms of the basis $e_1,e_2,e_3$ above one finds:
\be
\vec{\mu}  =  \left(\frac{r^2}{6}(1-y)(\cos\theta-1),~
\frac{r^2}{6}(1-y)\cos\theta-\frac{r^2}{2}l\ell y,~ \ell r^2 y\right)~. \ee
Notice that this involves the generically irrational parameter
$\ell $.

We will now describe the image of $\vec\mu$, and check that it is
given by a good convex rational polyhedral cone in $\R^3$,
as predicted by the results of \cite{FT, L}. First, note that the
edges of the cone can be identified by fixing any non--zero value
of $r$, say $r=1$, and then finding the submanifolds which are fixed under some $
\T^2\subset \T^3$ action. Indeed, the edges of the cone, which
generate it, are precisely the images of submanifolds in $C(Y^{p,q})$ over
which some two--torus collapses. There are four such submanifolds at $r=1$,
given by the north ($N$) and south ($S$) poles of
the base and fibre two--spheres: these are all copies of a circle --
specifically,
the fibre over the corresponding point on the base $B$.
We denote the subspaces as follows: $NN=\{y=y_2, \theta=0\}$,
$NS=\{y=y_2, \theta=\pi\}$,  $SN=\{y=y_1, \theta=0\}$,
$SS=\{y=y_1, \theta=\pi\}$. Then, using the useful
relations\footnote{One can derive these using the the explicit form for the
periods $P_i$ of Section 2, after using the cubic equation (\ref{thecubic}).}
\bea 1 - y_1 & = & - 3\ell k y_1 \nonumber\\
1-y_2 & = & 3 \ell l y_2~, \label{usfrel}
\eea
we find (at $r=1$)
\begin{eqnarray}
\vec\mu(NN) & = & \ell y_2~(0 ,~0 ,~1) \nonumber\\
\vec\mu(NS) & = & \ell y_2~(-l,~-l ,~1)\nonumber\\
\vec\mu(SN) & = & \ell y_1~(0,~-p ,~1)\nonumber\\
\vec\mu(SS) & = & \ell y_1~(k ,~q ,~1) ~.
\label{immap}
\end{eqnarray}
Note that the irrational parameter $\ell$ has factored out and the
vectors in (\ref{immap}) represent four lines
which are spanned as $r$ varies from $0$ to infinity. Noting that
$y_1<0$ and $y_2>0$ it is then
easy to verify that these are the edges of a four--faceted polyhedral
cone in $\R^3$ generated by:
\bea u_1= [0,p,-1], \quad u_2=[-k,-q,-1], \quad u_3=[0,0,1], \quad
u_4 = [-l,-l,1] 
\eea
with outward--pointing primitive normals:
\bea v_1 = [1,0,0],\quad v_2=[1,-2,-l],~\quad v_3 = [1,-1,-p],\quad
v_4 = [1,-1,0]~.
\eea
As described above, these normals characterise codimension two fixed
point sets in $C(Y^{p,q})$ over
which a circle of the three--torus shrinks to zero size.
The corresponding linear combination of Killing vectors
in $[\de/ \de \phi, \de / \de  \psi, \de / \de \gamma]$ should
then have vanishing norm when restricted to the pre--image of the facet.
Indeed, using the metric (\ref{tinky}) it is straightforward to verify
that the four Killing vectors
\bea V_1=\frac{\de}{\de \phi}+ \frac{\de}{\de \psi},\quad V_2=-
\frac{\de}{\de \phi}+ \frac{\de}{\de \psi},\quad
V_3=\frac{\de}{\de \psi}-\frac{k}{2}  \frac{\de}{\de \gamma},\quad
V_4=\frac{\de}{\de \psi}+\frac{l}{2}  \frac{\de}{\de \gamma} \label{vecs}\eea
vanish on the submanifolds given by $\theta=0$, $\theta=\pi$, $y=y_1$, and
$y=y_2$ respectively. Note that the normals obtained with the moment map use only the
{\em symplectic} structure of the manifolds, whereas the norms of
the Killing vectors are computed using the {\em metrics}.

Let us now make the following observations about the normal vectors
$v_1,\ldots,v_4$:
\begin{itemize}
\item $\{v_1,\ldots,v_4\}$ span $\Z^3$ over $\Z$. Indeed it
is trivial to see that $<v_1,v_4>=<E_1,E_2>$. The direction $<E_3>$ is then
obtained as a linear combination of  $v_1,v_2,v_3,v_4$. 
Indeed, since $\mathrm{hcf}(l,p)=1$ by Euclid's algorithm
there are integers $a,b\in \Z$ such that $al+bp=1$.

\item For each of the four edge vectors $u_i$, $i= 1,\ldots,4$, the
corresponding two normal vectors $v_{i_1}$, $v_{i_2}$, $i_1\neq i_2 \in \{1234\}$
with $u_i\cdot v_{i_1}=u_i\cdot v_{i_2}=0$ satisfy
\be
\{a_1v_{i_1}+a_2v_{i_2}\mid a_1,a_2\in \R\}\cap \Z^3 =
\{a_1v_{i_1}+a_2v_{i_2}\mid a_1,a_2 \in \Z\}~.\ee
\end{itemize}
The second condition is precisely the condition that the cone is
\emph{good}, in the sense of reference \cite{L}. Indeed, this must be true since
in \cite{L} it is shown that the image of a symplectic toric cone
under its moment map is always a good rational polyhedral cone. The
first property does not generically hold, but is special to the geometries
we are considering. As we will see later, it is related to the fact that
the Sasaki--Einstein manifolds we began with are simply--connected.

It will be useful to know the topology of the codimension two
submanifolds. Let us denote them as $F_i$, where $i=1,\ldots,4$,
respectively. Explicitly we have $F_1=\{\theta=0\}$,
$F_2=\{\theta=\pi\}$, $F_3=\{y=y_1\}$, $F_4=\{y=y_2\}$. If we project
out the $\gamma$ direction, these are all copies of $S^2$.  The first
two, $F_1/U(1)$, $F_2/U(1)$, are the two fibres of $B=S^2\hookrightarrow
S^2$ over the north and south poles of the
base $S^2$, and so are representatives of the cycle $C_1$. The third and
fourth, $F_3/U(1)$, $F_4/U(1)$, are the sections of the $S^2$
bundle at the south and north poles of the fibre $S^2$,
respectively\footnote{These were denoted $S_1$ and $S_2$ in \cite{paper2}.}.
Since the $\gamma$
direction describes a principle $U(1)$ bundle over each of these
spheres, the total spaces $F_i$ will be Lens spaces $L(1,m)\cong S^3/\Z_m$ for various
values of $m\in \mathbb{Z}$. To see which Lens spaces one has, one can
simply integrate the curvature two--form $\ell^{-1} \diff A$ over
$F_i/U(1)$ for each $i=1,\ldots,4$. One finds
\be
F_1 \cong F_2 \cong S^3/\Z_p, \quad F_3 \cong S^3/\Z_k, \quad F_4
\cong S^3/\Z_l~.
\ee
Thus the facets of the polyhedral cone lift to cones over the above
four Lens spaces. The latter two are calibrated submanifolds, as we 
saw in Section 2.

\section{Gauged linear sigma models}

In this section we begin by giving a brief review of gauged linear
sigma models \cite{wittenLSM}. We then move on to describe
Delzant's construction \cite{D} which from a polytope $\Delta$ constructs a
gauged linear sigma model whose vacuum manifold is precisely the
symplectic toric manifold corresponding to $\Delta$. The
construction also goes through for cones, provided one starts
with a good convex rational polyhedral cone \cite{L}.
We then use this method to construct the sigma model
for the cone $C(Y^{p,q})$. Using this approach, 
turning on Fayet--Iliopoulos parameters in the linear
sigma model one (partially) resolves the conical singuarity. As a
check on our result, we explicitly show how one can recover the
topology and group action on $Y^{p,q}$ from the linear sigma model
description, thus closing the loop of arguments. This is
summarised below:
\be
C(Y^{p,q})\stackrel{\mathrm{moment\ map}}{\longrightarrow} \mathrm{polyhedral \ cone} \subset \R^3
\stackrel{\mathrm{Delzant}}{\longrightarrow} \mathrm{\ linear\ sigma\ model}
\stackrel{\mathrm{vacuum}}{\longrightarrow} C(Y^{p,q}) \nonumber \ee

\subsection{A brief review}

Let $z_1,\ldots,z_d$ denote complex coordinates on $\C^d$. In physics terms these will be the
lowest components of chiral superfields $\Phi_i$, $i=1,\ldots,d$. We may specify an action of the
group $\T^r\cong U(1)^r$ on $\C^d$ by giving the integral charge matrix $Q=\{Q^i_{a}\mid i=1,\ldots,d ; \ a=1,\ldots,r\}$; here
the $a$th copy of $U(1)$ acts on $\C^d$ as
\be
(z_1,\ldots,z_d)\rightarrow (\lambda^{Q^1_a}z_1,\ldots,\lambda^{Q^d_{a}}z_d)\ee
where $\lambda\in U(1)$. We may then
perform the \emph{K\"ahler quotient} $X=\C^d//U(1)^r$ by imposing the $r$ constraints
\be
\sum_{i=1}^d Q^i_a |z_i|^2 = t_a\quad\quad a=1,\ldots,r \label{Dterms}
\ee
where $t_a$ are constants, and then quotienting out by $U(1)^r$. The resulting space has complex dimension
$n=d-r$ and inherits a K\"ahler structure, and thus also a symplectic structure, from that of $\C^d$. In physics terms,
the constraints (\ref{Dterms})
correspond to setting the $D$--terms of the gauged linear sigma model to zero to give the vacuum, where $t_a$ are
Fayet--Iliopoulos parameters -- one for each
$U(1)$ factor. The quotient by $\T^r$ then removes the gauge degrees of freedom. Thus the K\"ahler quotient
of the gauged linear sigma model precisely describes the classical vacuum of the theory. Note that the K\"ahler class of the
quotient $X$ depends linearly on the FI parameters $t_a$, and moreover even the topology of the quotient will depend
on these. Also observe that, setting all $t_a=0$, the resulting
quotient will be a cone. One sees this by noting that $z_i\rightarrow \nu z_i$, $i=1,\ldots,d$ is a symmetry in this case,
where $\nu\in \R^+$. The conical singularity is located at $z_i=0$, $i=1,\ldots,d$.

It is also an important fact that $c_1(X)=0$ is equivalent to the statement that the sum of the $U(1)$ charges
is zero for each $U(1)$ factor. Thus
\be
\sum_{i=1}^d Q^i_a = 0\quad\quad a=1,\ldots,r~.\ee
This latter fact ensures also that the one--loop beta function is zero. 
The sigma model is then Calabi--Yau, although
note that the metric induced by the K\"ahler quotient is \emph{not} Ricci--flat.

\subsection{Delzant's construction: from polytopes to gauged linear sigma-models}

Let us first suppose we have a Delzant polytope $\Delta$ which is the image of some 
compact symplectic toric
manifold $M$ under its associated moment map. One can reconstruct $M$ from $\Delta$ as
follows. Let $v_i\in \Z^n$, $i=1,\ldots,d$, denote the outward pointing primitive normal vectors to
the facets of $\Delta$. For some $\lambda_i\in\R$ we may then write
\be
\Delta = \left\{x\in \R^n\mid <x, v_i> \leq \lambda_i, i=1,\ldots,d\right\}~.
\ee
Consider now the linear map $\pi: \R^d \rightarrow \R^n$ which maps the standard basis
 vectors $E_i$ of $\R^d$
to $v_i$. Thus $\pi(E_i)=v_i$ for each $i=1,\ldots,d$. From the Delzant properties of $\Delta$ one easily
sees that this map is surjective. The kernel has dimension $r=d-n$, and defines a
corresponding torus $\T^r\subset \T^d$.
Now take $\C^d$ with its usual action by $\T^d$, and consider the moment map where 
we take the Fayet--Illiopoulos parameters to be $t_i=\lambda_i$. From the induced action 
by $\T^r\subset \T^d$ above,
we get an induced moment map for the $\T^r$ action. One may now take the symplectic 
reduction $\C^d//\T^r$, which is a
symplectic manifold of complex dimension
$d-r=d-(d-n)=n$. Moreover, this quotient also inherits an action of $\T^n=\T^d/\T^r$ from that of
$\C^d$ and is thus toric. In fact, it is not difficult to see that the image of $\C^d//\T^r$
under its moment map, associated to $\T^n$, is just $\Delta$. This is Delzant's construction \cite{D}.

As a completely trivial example, consider the two--sphere $S^2$ with
canonical $U(1)$ action which rotates about the equator. The image of
the moment map is just a line segment, with length proportional to the
volume of the two--sphere. The outward pointing normal one--vectors are
$v_1=1$, $v_2=-1$. The kernel of the map $\pi:E_i\mapsto v_i$ is
thus $(1,1)$, whence we see that $S^2$ is the symplectic reduction
of $\C^2$ by $U(1)$ with charges $(1,1)$ -- the $U(1)$ quotient is just 
the Hopf map $S^3\rightarrow S^2$.

There is a corresponding construction for compact symplectic toric
orbifolds, which is a generalisation that takes
into account that the normals may no longer form a $\Z$--basis for
$\Z^n$. This introduces finite subgroups $\Gamma$ which become
local orbifold groups in the symplectic quotient \cite{LT}.

\subsubsection*{A Delzant construction for cones}

Recently a Delzant theorem has been proven for symplectic toric cones
\cite{L}. The language used is largely that of contact
geometry -- recall that a metric cone over a contact manifold is precisely a symplectic
cone, and vice versa. The essential point is that the convex
rational polyhedral cone one starts with must be \emph{good}.
This ensures that the symplectic quotient is smooth.

Since the moment cones $\mu(C(Y^{p,q}))$ are all good cones, we may apply the theorem of
\cite{L}: one simply applies Delzant's construction, as in the compact case, 
and sets all the Fayet--Iliopoulos parameters to zero.
Thus recall that the outward pointing primitive normal vectors were found to be
\bea v_1 = [1,0,0],\quad v_2=[1,-2,-l],~\quad v_3 = [1,-1,-p],\quad
v_4 = [1,-1,0]~. \eea
By inspection the kernel is $(p,p,-l,-k)$. Thus the Delzant theorem for
cones gives

\begin{itemize}
\item {\sl $U(1)$ gauged linear sigma--model on
$\C^4$ with charge vector $Q=(p,p,-l,-k)$.}
\end{itemize}

As a preliminary check this this is indeed correct, notice that the charges
sum to zero: $p+p-l-k=0$, since $k=p+q$, $l=p-q$. It follows that the vacuum manifold $X$ of
this gauged linear sigma model is topologically
Calabi--Yau, $c_1(X)=0$, just as expected. Moreover, by turning on the
Fayet--Iliopoulos parameter $t$ for the $U(1)$ gauge field we will
obtain orbifold resolutions of the cone.

As interesting degenerate cases, consider $p=1, q=0$. This is the
(resolved) conifold, which recall is the gauged linear sigma model
on $\C^4$ with charges $Q=(1,1,-1,-1)$. Another important case is
$p=q=1$. This yields $\C$ times the linear sigma model on $\C^3$
with charges $(1,1,-2)$. The latter is $\mathcal{O}_{\cp^1}(-2)$. Taking $t=0$ shrinks the $\cp^1$ to zero size, yielding
the orbifold $\C^2/\Z_2$, which is also the $A_1$ singularity.
Thus the cone is $\C \times (\C^2/\Z_2)$. This has $\mathcal{N}=2$
rather than $\mathcal{N}=1$ supersymmetry. The horizons of these
two spaces are thus $T^{1,1}$ and $S^5/\Z_2$. 

If one formally takes  $p=q\neq 1$ and $q=0$, one obtains $\Z_p$
quotients of the cases above. In particular these will correspond to orbifolds
$(\C^2/\Z_2\times \C)/\Z_p$ and (conifold)$/\Z_p$ respectively. It 
is interesting to note that these are consistent with the limiting volumes
(\ref{limitvolumes}), although the metrics $Y^{p,q}$ are not valid in these limits.

We can now use the results of \cite{Ltop} to perform further non--trivial checks.
According to Theorem 1.1 of Ref. \cite{Ltop} we have the following
general topological facts about the base $Y$ of the symplectic toric cone
$C(Y)$ we began with (provided it is of Reeb type):
\begin{itemize}
\item $\pi_1(Y)\cong \Z^n/<v_i>$, is a finite abelian group. Recall that
$n=\mathrm{dim}(\T^n)$ is the complex dimension of the cone $C(Y)$.

\item $\pi_2(Y)$ is a free abelian group of rank $d-n$, where
$d$ is the number of facets of the moment cone.
\end{itemize}
We may now verify that these are indeed true for our examples $Y^{p,q}$
and their moment cones.
In particular, for our polyhedral cones recall that the $\{v_i\}$ spanned
$\Z^3$ over $\Z$, and thus $\pi_1(Y^{p,q})$ is trivial, in agreement with
the fact that $Y^{p,q}\cong S^2 \times S^3$ for all $p,q$. Moreover, we
may now relax the condition that $\mathrm{hcf}(p,q)=1$. From the Gysin sequence
for the $U(1)$ fibration corresponding to $\de/\de\gamma$,
as in the appendix of \cite{paper2}, one sees that $\pi_1(Y^{p,q})\cong \Z_h$
where $h\equiv \mathrm{hcf}(p,q)$. Since now $\mathrm{hcf}(l,p)=\mathrm{hcf}(p,q)= h$
Lerman's theorem says that $\pi_1(Y^{p,q})\cong \Z_h$, in agreement with the
Gysin sequence calculation.

For the second point in the theorem, since there are four normals, we also learn that
$\pi_2(Y^{p,q})\cong \Z$, again in perfect agreement with the topology we started with.

\subsection{The topology of the vacuum}

In this subsection we verify that one can recover the topology of, as well as the action of the isometry group on, $Y^{p,q}$ correctly as the
boundary, or horizon, of the linear sigma model
$(p,p,-l,-k)$. Of course, this is guaranteed by the Delzant theorem of
\cite{L}.
Nevertheless, it is interesting to analyse the relation explicitly, since this
sheds considerable light on the geometry and topology. 

Since this
``hands on'' approach is rather technical, the reader might well
omit the remainder of this section. However, we will need the
relation (\ref{relvec}) between vectors on $\C^4$ and $Y^{p,q}$ in
the next section. This section also constitutes a direct proof of the 
equivalence of the gauged linear sigma models with the Calabi--Yau cones, 
without using any theorems.

\subsubsection*{A direct analysis of the topology}

The point of this subsection is to show that the K\"ahler quotient 
$\C^4//U(1)$ is topologically the same as $C(Y^{p,q})$. This is far from 
obvious, but is nevertheless guaranteed by the general theorems we have 
used thus far.

At $z_3=z_4=0$ we have a finite sized
$\cp^1$, of size $t/p$, where recall that $t$ is the FI parameter. We may thus introduce
gauge invariant coordinates $z=z_1/z_2$, $z^{\prime}=z_2/z_1$ which
cover the open subsets
$U_2,U_1\subset \cp^1$ where $U_i=\{z_i\neq0, z_3 = z_4 = 0\}\subset
\cp^1$. On the overlap $U_2\cap U_1$
we have $z=1/z^{\prime}$, thus making the Riemann sphere. However, for $p>1$
this $\cp^1$ is a locus of $\Z_p$ orbifold singularities in the K\"ahler quotient. Indeed, the
subgroup $\Z_p\subset U(1)$ stabilises the subspace $(z_1,z_2,0,0)$ of
$\C^4$. The fact that we have a non--trivial isotropy subgroup means
that this will descend to a locus
of $\Z_p$ orbifold singularities in the quotient space. To analyse
this singularity, consider, for
example, the subspace given by $z_1=0$. Using a gauge transformation
we may set $z_2$ to be real
and positive, which is thus the north pole of the base $\cp^1=S^2$.
The action of $\Z_p$ on $(z_3,z_4)$ is generated by
$(z_3,z_4)\rightarrow (z_3\omega_p^{-l}, z_4\omega_p^{-k})$ where $\omega_p =
e^{2\pi i/p}$ generates $\Z_p$. Note that this is equivalent to the
anti--diagonal action
$(z_3,z_4)\rightarrow (z_3 \omega_p^{q},
z_4\omega_p^{-q})$. Thus if $U(1)_A\subset SU(2)\subset U(2)$ acts on
$\C^2$ in the usual
way, we have that the generator $\omega_p$ of $\Z_p$ embeds in
$U(1)_A$ as $\omega_p^q$. Notice that $|z_2|^2\geq t/p$ and that, for
fixed $|z_2|^2>t/p$ the
$D$--term imposes that the coordinates $z_3,z_4$ define an ellipsoid,
which topologically is
$S^3$ modulo the $\Z_p$ action just discussed. Since $q$ is prime to
$p$ this is the Lens space $L(1,p)$. At $|z_2|^2=t/p$ we have
$z_3=z_4=0$ and the Lens space collapses. Thus the subspace $z_1=0$ is
a copy of an $A_{p-1}$ singularity. Performing the quotient
of the Lens space ``at infinity'' by $U(1)_A$ then
gives a two--sphere, the
map being the $p$th power of the anti--Hopf
map\footnote{Note the distinction here with the diagonal subgroup
  $U(1)_D$ of $U(2)$. Quotienting
by this is the Hopf map, and moreover since this is a normal
subgroup the quotient is also group $U(2)/U(1)_D\cong SO(3)$. This
$SO(3)$ thus acts naturally on the projected space.}.

Clearly the same picture holds at all points in $\cp^1$, not just at $z_1=0$,
as $SO(3)$ acts as a symmetry. It follows that we have an $A_{p-1}$ fibration over
 this $\cp^1$, which thus has a boundary which is a Lens space bundle
over $\cp^1$. In fact such a bundle structure of the metrics $Y^{p,q}$ was already
noted in reference \cite{paper2}.
We may then quotient the boundary by $U(1)_A$ to obtain a space $\hat{B}$ that
will be an $S^2$ bundle over the
base $\cp^1=S^2$. To see what this bundle is we may introduce
coordinates as follows. Suppose
$z_2\neq0$, giving the patch $U_2$ on the
base $\cp^1$ with coordinate $z=z_1/z_2$. In order to effectively go
to the boundary of our space,
we may set $l|z_3|^2+k|z_4|^2=\mathrm{constant}>0$.
In particular,
we cannot have both $z_3$ and $z_4$ zero. Suppose then that
$z_3\neq0$. We may now introduce the
additional coordinate
$x_2=\bar{z}_4/z_3z_2^2$ on the fibre. This
is invariant under both
the original $U(1)$ action -- the key point being that $k+l=2p$ -- as
well as $U(1)_A$ under which the
fields have charges
$(0,0,1,-1)$. Similarly over $U_1$ we have coordinate $x_1 =
\bar{z}_4/z_3z_1^2$. The union of these two subspaces thus describes the
bundle $\mathcal{O}_{\cp^1}(2)$. However, note that, due to
the presence of the
$\bar{z}_4$s, the complex structure here is
\emph{not} inherited from the complex structure of $\C^4$ we started
with. Since we are only
interested in topology and group actions, this
fact will not be important for the present discussion. Similarly, for $z_4\neq0$ one has coordinates
$w_2=z_3z_2^2/\bar{z}_4$
and $w_1=z_3z_1^2/\bar{z}_4$. This describes
$\mathcal{O}_{\cp^1}(-2)$. The intersection
of these subspaces results in the gluing
of the two $\C$ fibres together to create a Riemann sphere $S^2$ bundle over
$\cp^1=S^2$ -- for example, $x_2=1/w_2$ on the overlap with
$z_2,z_3,z_4\neq0$. Thus we obtain
precisely the same description
as the manifold $B$ discussed earlier: $\hat{B}\cong B$.

Topologically, the manifold $\hat{B}$ just described is the same
thing as $\mathbb{P}(\mathcal{O}\oplus \mathcal{O}(-2))$ which is
the second Hirzebruch surface $\mathbb{F}_2$. However, due to the
$\bar{z}_4$s, the complex structure is not that inherited from $\C^4$. Indeed,
if one replaces $\bar{z}_4$ by $z_4$ in the above coordinates, one
precisely gets $\mathbb{F}_2$, as one can see by analysing the
linear sigma model for this manifold\footnote{ This is a $U(1)^2$
model on $\C^4$ with charges $Q_1=(1,1,2,0)$, $Q_2=(0,0,1,1)$.}.
The fibre $S^2$ is thus perhaps best described as
$\overline{\cp}^1$. Moreover, as explained in \cite{paper2}, as a
real manifold $\mathbb{F}_2$ is actually just a product space $S^2
\times S^2$ {\it i.e.} the bundle is trivial.

It remains to compute the twisting of $U(1)_A$ over this base $B$,
which as we have just seen
is naturally described as an $S^2$ bundle over $S^2$ with twist 2.
Over the fibre
$S^2$, sitting at some point on the base $S^2=\cp^1$, the twisting of the
$U(1)$ is $p$, as is clear
from the above discussion since the
fibre sphere descended from the Lens space
$L(1,p)\cong S^3/\Z_p$. We now compute the $U(1)$ twisting over the
copies of $S^2$ at the south and north poles of the fibre $S^2$ --
these are two sections of
the $S^2$ bundle. Call them $S_1$ and
$S_2$, respectively, as in \cite{paper2}. These are given by
$z_3=0$, $z_4=0$, respectively, which give linear sigma models on $\C^3$
with weights $(p,p,-k)$, $(p,p,-l)$, respectively. The boundaries of
these two spaces are Lens
spaces $L(1,k)$, $L(1,l)$. To see this,
note that $S^1/\Z_p\cong S^1$. Thus the boundaries are $S^1$ bundles
over $S^2$. The twisting
in each case is easily seen to be
$k$ and $l$, respectively.

We may now relate this to our earlier discussion. Recall that the
canonical generators
$C_1,C_2$ of the second homology of
$S^2 \times S^2$ are
related to the copies $S_1, S_2$ of $S^2$ at the south and north poles of the fibre $S^2$ by
\bea
2C_1 &=& S_1 - S_2\\
2C_2 &=& S_1 + S_2~.
\eea
We have just seen that the twisting over $S_1$ and $-S_2$ is $k$ and
$l$, respectively. This gives
the Chern numbers over
$C_1$ and $C_2$ to be $(k+l)/2=p$, and $(k-l)/2=q$, respectively. We thus precisely reproduce the topology
of $Y^{p,q}$ described in section 2. Moreover, the (not quite effectively acting)
isometry group of the Sasaki--Einstein metrics is
$SU(2)\times U(1)^2$. The K\"ahler quotient above also has
this isometry group --
this is just the subgroup of $U(4)$ that commutes with the original
$U(1)$ action.

\subsubsection*{Relation between Killing vector fields}

It is also now interesting to examine the codimension two fixed point sets of 
the linear
sigma model $(p,p,-l,-k)$ directly, and compare with our polyhedral
cone for $C(Y^{p,q})$. Thus we now set $t=0$. The codimension two fixed point sets are easily
found: they are at $z_i=0$, for each $i=1,\ldots, 4$. Indeed, from our
above discussion of the topology of the vacuum, these are precisely cones over the Lens spaces
$S^3/\Z_p$, $S^3/\Z_p$, $S^3/\Z_k$, $S^3/\Z_l$, respectively. In terms
of $Y^{p,q}$, these are the submanifolds $F_i$, $i=1,\ldots,4$, respectively. In
particular, note that $F_3/U(1) \cong S_1$, $F_4/U(1)\cong S_2$. Thus
we see explicitly that the topology of the subspaces $\{z_i=0\}$ are the
same as $C(F_i)$, respectively.

The relation between the Killing vectors is also easy to make
explicit. Let us denote $\de/\de \theta_i$ as the $U(1)$ that rotates
the coordinate $z_i$. Thus $\de/\de\theta_i=0$ defines the codimension
two submanifolds $z_i=0$. We find
\begin{eqnarray}
2\frac{\de}{\de\phi} & = &
\frac{\de}{\de\theta_1}-\frac{\de}{\de\theta_2} \nonumber\\
2p\frac{\de}{\de\psi} & = &
l\frac{\de}{\de\theta_3}+k\frac{\de}{\de\theta_4} \nonumber\\
p\frac{\de}{\de\gamma} & = &
-\frac{\de}{\de\theta_3}+\frac{\de}{\de\theta_4}~.\end{eqnarray}
These require some explanation. We denote the weights of the
$\partial/\partial\theta_i$ as a row vector for convenience. Thus consider
$(1,-1,0,0)$. For $t>0$ this precisely rotates the subspace
$z_3=z_4=0$, which is a copy of $\cp^1$ of size $t/p$, with weight \emph{two}.
Hence we identify this $U(1)$ with
$2\de/\de\phi$. Also, by construction, the $\de/\de\gamma$ direction is proportional
to $(0,0,1,-1)$ which recall we denoted $U(1)_A$. However, the orbits
of the vector $(0,0,-1,1)$ actually wind $p$ times around the circle
fibre: recall the projection of this $U(1)$ was the $p$th power of the
anti--Hopf map. Hence this is $p\de/\de\gamma$. Finally, note that
$\de/\de\psi$ rotates the fibre $S^2$ with weight one and does not act
on the base $S^2$. This determines
the final vector, as one can see by analysing the action on the coordinates
$x_1,x_2,w_1,w_2$ introduced above.

To make contact with the normal vectors discussed earlier, one must
note that the Killing vector given by $(p,p,-l,-k)$ acts trivially on
the vacuum, by construction. Thus $(p,-p,0,0)$ is equivalent to both
$(2p, 0, -l,-k)$ and $(0,-2p,l,k)$. Thus we compute
\begin{eqnarray}
\frac{\de}{\de\phi}+\frac{\de}{\de\psi} & = & \frac{\de}{\de\theta_1}
\nonumber\\
-\frac{\de}{\de\phi}+\frac{\de}{\de\psi} & = &
\frac{\de}{\de\theta_2} \nonumber\\
\frac{\de}{\de\psi}-\frac{k}{2}\frac{\de}{\de\gamma} & = &
\frac{\de}{\de\theta_3} \nonumber\\
\frac{\de}{\de\psi}+\frac{l}{2}\frac{\de}{\de\gamma} & = &
\frac{\de}{\de\theta_4} \label{relvec}\end{eqnarray}
in perfect agreement with our earlier results: the vectors on the left hand 
side are precisely the Killing vectors (\ref{vecs}) which fixed codimension 
two submanifolds of $Y^{p,q}$. In particular, this means that the polyhedral
cones for $C(Y^{p,q})$ and the linear sigma model with weights
$(p,p,-l,-k)$ are identical, and thus they are completely equivalent
as symplectic toric cones {\it i.e.} they are \emph{equivariantly
symplectomorphic}. We have shown this directly in this subsection,
without appealing to any theorems.

\section{Toric Gorenstein canonical singularities}

In this section we make contact with reference \cite{MP}
by explaining the relation of the Calabi--Yau
gauged linear sigma model $(p,p,-l,-k)$ to so--called toric Gorenstein
canonical singularities.

The data required to define a \emph{toric Gorenstein canonical
singularity} of complex dimension $n$ is a convex polygon on
$\R^{n-1}$, all of whose vertices have integer coordinates. Given
any such polygon one can reconstruct the toric singularity, as
well as all of its toric crepant resolutions, as follows. Let
$\{\mathcal{V}_{i}\mid i=1,\ldots, d\}$ denote all vectors in $\R^{n-1}$
with integer coordinates and with the property that they lie
within, or on the boundary of, the polygon. Marking these points
gives the toric diagram $\mathcal{D}$. Consider now the set of all
linear relations among these vectors
\be \sum_{i=1}^d Q^{i}_a \mathcal{V}_{i} = 0\ee
with integer coefficients $Q^{i}_a$ satisfying
\be \sum_{i=1}^d Q^{i}_a = 0\ee
for each $a=1,\ldots r$, where $a$ labels the set of such linear
relations. Clearly $r=d-n$. One now uses the matrix $Q^i_a$ as the
charges of a linear sigma model on $\C^d$ with gauge group
$U(1)^r$. This is essentially a Delzant construction. The K\"ahler quotient $X=\C^d//U(1)^r$ has complex
dimension $n=d-r$. Setting all FI parameters to zero gives the
toric singularity. Moreover, by turning on the FI parameters one
obtains (partial) resolutions of the singularity -- special values
of the FI parameters will give rise to more singular spaces than
the generic values. By including all the interior points $\mathcal{V}_i$ of
the polygon, we have ensured that the linear sigma model
reproduces all the toric crepant resolutions of the singularity.
The sizes of the blow--ups are controlled by the FI parameters.
However, this is not usually a very economical way of constructing
the singularity -- the minimal presentation, meaning the smallest
possible $d$ and thus least number of chiral superfields, arises by
using only the vertices of the polygon\footnote{If these vectors
do not span $\Z^{n-1}$ over $\Z$ one must in addition
quotient the K\"ahler quotient by the finite group $\Z^{n-1}/<\mathcal{V}_i>$
to correctly reproduce the singularity -- this follows from our 
general discussion in Section 3.}.

The toric diagram for the Calabi--Yau cone on $Y^{p,q}$ can be
obtained as follows. Recall that the image of the
moment map for $C(Y^{p,q})$ is a four--faceted polyhedral cone with
primitive outward pointing normals
\bea v_1 =[1,0,0], \quad v_2=[1,-2,-l], \quad v_3=[1,-1,-p], \quad v_4=[1,-1,0]~.\eea
Notice that these vectors lie in the plane at $e_1=1$.
Indeed, the normals belong to a plane in $\R^2$ precisely when the
linear sigma model is Calabi--Yau. Thus we may project onto the $e_1=1$ plane
to obtain vectors
\bea [0,0], \quad [-2,-l], \quad [-1,-p], \quad [-1,0]~.\eea
We now shift the origin by $[1,0]$ and then make the $SL(2;\Z)$ transformation
\begin{eqnarray}\left(
\begin{array}{cc}
l-1 & -1 \\
l  & -1 \\
\end{array}\right)\end{eqnarray}
to obtain vectors
\bea \mathcal{V}_1 = [l-1,l], \quad \mathcal{V}_2 = [1,0], \quad \mathcal{V}_3 = [p,p], \quad \mathcal{V}_4 = [0,0]\eea
respectively. This is a minimal presentation of the singularity.
The pictures below display some examples with low values of
$p$.
\begin{figure}[!h]
\vspace{5mm}
\begin{center}
\epsfig{file=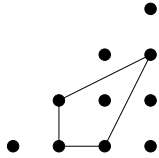,width=2cm,height=2cm}\\
\end{center}
\caption{Toric diagram of $Y^{2,1}$ embedded in the orbifold
$\C^3/\Z_3\times \Z_3$.} \vspace{5mm}
\label{figreid21}
\end{figure}
It is interesting to note that the areas of these polygons are equal to
$p$, independently of $q$. Indeed, for fixed $p$, varying $q$ just slides
the vertex $\mathcal{V}_1$ up and down the hypotenuse of the triangle that
defines the orbifold $\C^3/\Z_{p+1}\times\Z_{p+1}$.
Note that for $(p,q)=(2,1)$ the toric diagram
is the same as that for the complex cone (canonical line bundle) over the first del
Pezzo surface, as we discuss in detail in the following section.
\begin{figure}[!h]
\vspace{5mm}
\begin{center}
\epsfig{file=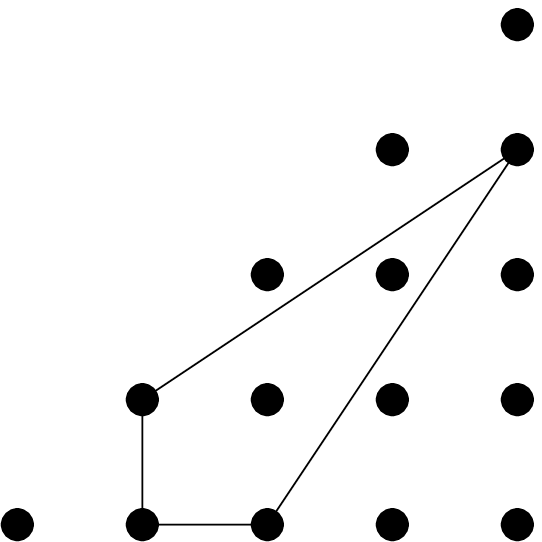,width=2.5cm,height=2.5cm}\hspace{4cm}
\epsfig{file=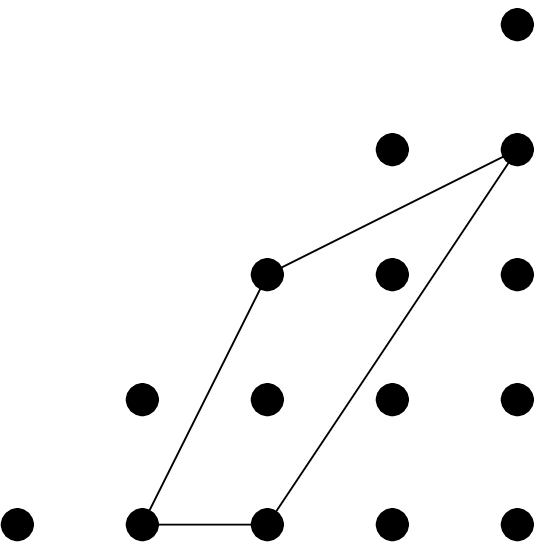,width=2.5cm,height=2.5cm}\\
\end{center}
\caption{Toric diagrams of $Y^{3,2}$ and $Y^{3,1}$
embedded in the orbifold $\C^3/\Z_4\times \Z_4$.} \vspace{5mm}
\label{figreid32}
\end{figure}

Let us also remark that the number of points inside the polygon is
precisely $p-1$. Each point corresponds to a normal vector to a plane in
$\R^3$.
The total number of Fayet--Iliopoulos
parameters (K\"ahler parameters) is $(4-3)+p-1=p$, and by varying these
one moves the planes in their normal directions so that they no longer
intersect the origin. By assigning
generic values one completely resolves the conical singularity.
Indeed, these parameters roughly control the size of $\cp^1$s. We thus learn
that the Calabi--Yau cone, where all FI parameters are set to zero,
has $p$ collapsed two--spheres. Turning on the FI parameter $t>0$ in
the linear sigma model $(p,p,-l,-k)$ partially resolves the singularity
to an $A_{p-1}$ singularity fibred over $\cp^1$, as discussed in the last
section. Indeed, an $A_{p-1}$ singularity can be completely resolved
by blowing up $(p-1)$ two--spheres -- the metric is the $p$--centered
Gibbons--Hawking metric. There are precisely $(p-1)$ FI parameters, giving
$1+(p-1)=p$ in total.

\section{The complex cone over $\F_1$}

As noted above, the toric diagram we have found for $Y^{2,1}$ is the
same as that for the complex cone over the first del Pezzo surface. We
will refer to the latter as $\F_1$ and its complex Calabi--Yau
cone as $C_{\C}(\F_1)$.
Here we elaborate on this point. In particular, it follows that
we will inherit a metric on $\F_1$ from
that on $Y^{2,1}$, and we will write this down explicitly. 
Of course this metric will not be K\"ahler--Einstein.

First we will use the toric data we have to deduce the Killing
vector field on $Y^{2,1}$ corresponding to the complex cone direction.
Adapting the metric to this direction, we shall indeed find a smooth
metric on $\F_1$.

We label the five vertices of the toric diagram, including the blow--up
mode corresponding to the interior point, as
\be
\mathcal{V}_1 = [0,1], \quad \mathcal{V}_2 = [1,0],\quad \mathcal{V}_3 = [2,2], \quad \mathcal{V}_4 = [0,0],
\quad \mathcal{V}_5 = [1,1]~.\ee
The last vector $\mathcal{V}_5$ is the additional blow--up vertex. A
possible basis for the two charge vectors is given by
\bea\label{cf1}
Q_1 & = & (1,1,0,-1,-1)\nn\\ 
Q_2 & = & (0,0,1,\ 1,-2)~.\eea
We thus obtain a gauged linear sigma model on $\C^5$ with $U(1)^2$
gauge group. Let us for the moment drop the last entry in these vectors.
This gives a gauged linear sigma model on $\C^4$ with weights
\bea\label{f1}
\hat{Q}_1 & = & (1,1,0,-1)\nn\\ 
\hat{Q}_2 & = & (0,0,1,\ 1)~.
\eea
Let us take each quotient in turn. The first quotient yields
$\C\times [\mathcal{O}_{\cp^1}(-1)]$, since $(1,1,-1)$
is precisely $\mathcal{O}_{\cp^1}(-1)$. The former may also
be regarded as $\mathcal{O}_{\cp^1}(0)\oplus\mathcal{O}_{\cp^1}(-1)$.
The second row then projectivises this $\C^2=\C\oplus\C$ bundle.
This means one quotients each $\C^2$ fibre by the Hopf map $\C^2\setminus
\{0\}\rightarrow \cp^1$. The resulting space is the first Hirzebruch surface
\be
\F_1=\mathbb{P}(\mathcal{O}_{\cp^1}(0)\oplus\mathcal{O}_{\cp^1}(-1))~.
\ee
This is also the same thing as $\cp^2$ blown up at a point\footnote{In the toric
language, there is a nice way to understand this. In fact, it's
straightforward to compute the Delzant polytope for $\cp^2$: this
is an isosceles rectangular triangle. A toric blow--up is obtained by simply chopping
off a vertex to give a rectangular trapezoid.}. Indeed,
$\cp^2$ may be obtained by taking $\mathcal{O}(1)\rightarrow \cp^1$
and gluing to its boundary, which is topologically $S^3$, a ball in $\C^2$.
Blowing up the origin in $\C^2$ replaces it by a $\cp^1$, which
has local geometry $\mathcal{O}_{\cp^1}(-1)$. Equivalently one
can describe this blowing up process as taking a connected sum with
$\cp^2$ with reversed orientation: $\cp^2\# -\cp^2$.
We now have
two copies of $\cp^1$ in the resulting space. In fact it is easy
to see that these are
two sections of $\F_1$ -- this is precisely analogous
to the topological construction of $B$. Note however that $w_2(\F_1)\neq 0$
and thus this is not a spin manifold.

Adding back the fifth entry to the charge vectors (\ref{f1}) to give
(\ref{cf1}) then describes the
canonical bundle over $\F_1$ -- the charges sum to zero, meaning that
the vacuum $X$ (K\"ahler quotient) is topologically Calabi--Yau, $c_1(X)=0$.
This identifies the canonical bundle, or complex cone, over $\F_1$.

Consider now taking a different linear combination of charge vectors,
corresponding
to a change of basis for the $\T^2$ action. In particular, using an
$SL(2;\Z)$ transformation we may take
\bea
Q_1^{\prime} &=& (2,2,-1,-3,0)\nn\\ 
Q_2^{\prime} &= &(1,1,0,-1,-1)~.\eea
The first set of weights of course gives the gauged linear sigma model on
$\C^4$ given by $(2,2,-1,-3)=(p,p,-l,-k)$, together with a factor of
$\C$. We may now effectively gauge away the second $U(1)$. Indeed, this means
\be
\frac{\de}{\de\theta_5}=-\frac{\de}{\de\theta_4}+\frac{\de}{\de\theta_1}+\frac{\de}{\de\theta_2} = \frac{\de}{\de\psi}-\frac{1}{2}\frac{\de}{\de\gamma}~.\label{theta5}\ee
acting on the linear sigma model $(2,2,-1,-3)$ on $\C^4$, and
$Y^{2,1}$, respectively. Here we have used the relations (\ref{relvec}).
Note that
$\de/\de\theta_5$ precisely rotates the complex line fibre over $\F_1$.
One can check explicitly that this Killing vector field on $Y^{2,1}$ is
nowhere--vanishing. Indeed, its norm--squared is computed to be
\be
\left|\frac{\de}{\de\theta_5}\right|^2 = F(y)\equiv
\frac{q(y)}{9}+w(y)\left(f(y)-\tfrac{1}{2}\ell\right)^2\ee
which is strictly positive. Here
\be
f(y)=\frac{a-2y+y^2}{6(a-y^2)}
\ee
is the function appearing in the local
one--form $A$. Of course in this particular case $a$ and $\ell$
take specific values. One finds
\bea
a  =  \frac{1}{2}(1-\tfrac{\sqrt{13}}{16}) &\qquad\qquad &
\ell  =  \frac{1}{2\sqrt{13}-5}\nn\\
y_1  =  \frac{1}{8}(1-\sqrt{13})&\qquad\qquad &
y_2  =  \frac{1}{8}(7-\sqrt{13})~.\eea
Let us summarise the situation. We have found that the metric $Y^{2,1}$
is an explicit irregular Sasaki--Einstein metric on the horizon of the
complex cone $C_{\C}(\F_1)$ over $\F_1$, where the Killing vector
field (\ref{theta5}) rotates the complex cone direction. Crucially this
is \emph{not} the Reeb vector, whose generic orbits in fact don't close.
The quotient of the metric (\ref{tinky}) by the
$U(1)$ action generated by (\ref{theta5}) should be a metric on $\F_1$. We
will now explicitly compute this metric and verify that it is
indeed a smooth metric on $\F_1$.

In order to perform the $U(1)$ quotient of $Y^{2,1}$,
it is useful to
first rewrite the metric adapted to the Killing vector field
$\de/\de \theta_5$. Thus, let
us change coordinates:
\be
\psi=\theta_5,\quad \gamma = -\Omega/2 -\theta_5/2~\label{ct}.\ee
It is then straightforward to compute the following expression for the metric

\begin{equation}
\label{newmetric}
\begin{aligned}
   \diff s^2 &= \frac{1-y}{6}(\diff\theta^2+\sin^2\theta
      \diff\phi^2)+\frac{1}{w(y)q(y)}\diff y^2+
      \frac{w(y)q(y)\ell^2}{36 F(y)}(\diff\Omega+\cos\theta \diff\phi)^2 \\
      & \qquad \qquad
      + {F(y)}\left[\diff\theta_5 - C \right]^2
\end{aligned}
\end{equation}
where we have defined
\bea
C = \frac{1}{F(y)}\left[ w(y)(f(y)-\tfrac{\ell}{2})\tfrac{\ell}{2}\diff \Omega
+\left( \frac{q(y)}{9}+w(y)f(y)(f(y)-\tfrac{\ell}{2})\right)\cos\theta\diff \phi\right]~.
\eea
The quotient by $\de/\de\theta_5$ now simply gives the metric in the first
line of (\ref{newmetric}), which again looks like a bundle over a
base two--sphere.
Let us now analyze regularity of this metric.

First, notice that all the functions are positive semi--definite. So, as usual, one
has to worry only about the smoothness conditions where the function $q(y)$
vanishes, and then check that the resulting periodicities give a
well--defined bundle--metric.
Near such a zero $y_i$, the ``fibre metric'', {\it i.e.} the
metric at fixed $\theta,\phi$,  takes the form
\bea
\diff s^2 (\mathrm{fibre}) & \approx &\frac{1}{12|y_i||y-y_i|}
\diff y^2 + \frac{|y_i| |y-y_i|\ell^2}{3F(y_i)} \diff \Omega^2~.
\eea
Now, crucially, the following relations are true for any $(p,q)$:
\bea
F(y_1) = (k-1)^2\ell^2y_1^2 \qquad \qquad F(y_2)= (l+1)^2\ell^2y_2^2~.
\label{crucial}\eea
Introducing $R=2|y-y_i|^{1/2}$ we find that
for $(k,l)=(3,1)$ -- and \emph{only} for these values -- the metric approaches
\bea
\diff s^2 (\mathrm{fibre}) & \approx & \frac{1}{12|y_i|} \left(\diff R^2 + \frac{1}{4}R^2\diff \Omega^2\right)
\eea
near the two zeros. We therefore obtain a smooth metric on $\R^2$ in this
neighbourhood if and only if $\Omega$ has period $4\pi$. Indeed, one can see
that this is the induced period for $\Omega$ from the metric on $Y^{2,1}$
by examining the coordinate transformation
(\ref{ct}): since $\psi,\gamma$ and $\theta_5$ all have period $2\pi$
one can calculate the period of $\Omega$ from the Jabobian of the
coordinate transformation (\ref{ct}), which is $-1/2$. This indeed means
that $\Omega\sim\Omega+4\pi$ and moreover with this period we have that
for fixed $y$, $y_1<y<y_2$, the resulting space is a squashed $S^3$.
These are then the generic orbits under the action of the isometry group
$U(2)$ on this  manifold.
We thus obtain an $S^2$ bundle over $S^2$ with twist one, which is
topologically $\F_1$, just as expected.

Let us now label the two sections of $\F_1$ at $y=y_1$, $y=y_2$ as
$H$, $E$ respectively. These are the hyperplane class 
and exceptional divisor of del Pezzo one, respectively.
It is a simple exercise to compute the Chern numbers of the $U(1)$
principle bundle, with coordinate $\theta_5$, over these:
\bea\label{chern}
\int_{H} \frac{\diff C}{2\pi}  ~=~  3\qquad\qquad\quad
\int_{E} \frac{\diff C}{2\pi}  ~=~  1\eea
where, as ever, we have to use the cubic (\ref{thecubic}) and, in this particular case,
$k=3$, $l=1$. Equations (\ref{chern}) give precisely the Chern
numbers required so that
the complex cone (or complex line bundle)
defined by the $U(1)$ bundle associated to $\theta_5$
is indeed Calabi--Yau. To see this, notice that the normal bundles
of the two $\cp^1$s corresponding to $H, E$ inside
$\F_1$ are topologically $\mathcal{O}_{\cp^1}(1)$ and
$\mathcal{O}_{\cp^1}(-1)$, respectively, as is clear from
our discussion of $\F_1$ above. 
Thus $c_1(\F_1)$ restricted to the two cycles gives $1+2=3$ and $-1+2=1$,
respectively, where $2=c_1(TS^2)$. The Chern numbers
above for $-\diff C$ precisely cancel these in the total space of
the associated complex line bundle, thus giving a Calabi--Yau manifold.

As shown at the end of Section 2.2, 
the cones over the $U(1)$ bundles over $H$ and $E$ (which are the 
submanifolds $y=y_1$, $y=y_2$) are divisors in the Calabi--Yau cone. 
Equivalently, the \emph{complex} cones over the submanifolds $H$ and $E$ are divisors. Indeed, we already noted above the normal 
bundles to these submanifolds inside $\F_1$, which translate into self--intersection 
numbers $H\cdot H = 1$, $E\cdot E = -1$.

One can check that the metric on $\F_1$ is not Einstein. Thus, in particular
it is not diffeomorphic to the Page metric on $\F_1$ \cite{page}, although it is  
rather similar in form.

\section{New non--trivial AdS/CFT predictions}

In this final section we discuss features of the gauge theory duals 
of the Sasaki--Einstein manifolds $Y^{p,q}$, focusing in particular on 
$Y^{2,1}$ since a candidate dual is already known. 
In particular we may compare our geometrical results to 
the $a$--maximisation calculation\footnote{Note that in \cite{BB} the central 
charge of the $dP_2$ quiver gauge theory is also calculated, and 
found to be quadratic 
irrational.} presented in \cite{BB}. 
We find complete agreement with this field theory calculation, both  
for the central charge and for the $SU(2)_F$ singlet baryons of the theory.

Let us first remark that, given a toric Gorenstein canonical 
singularity, an algorithm for constructing\footnote{Note that, in earlier work,\
extending that of \cite{KW}, the quiver gauge theories associated to some toric 
singularities were worked out in \cite{dallagata1,Fabbri,dallagata2} 
without using these algorithms.} a 
quiver gauge theory that has the singularity as its Higgs branch 
has been developed in \cite{hanany, hanany2} and subsequent works by these authors. This relies on the fact that any such singularity may be 
obtained by partial resolution of the orbifold $\C^3/\Z_{p+ 1}\times \Z_{p+1}$, 
and the field theory for the latter is known. In practise the 
algorithm requires a computer, even for relatively small $p$. However, 
the simple analytic expressions found in this paper suggest 
that  all theories can be treated simultaneously. Indeed 
it is tempting to speculate that some members of the family could be related
by deformations or connected via RG--flows. In particular, we can anticipate 
that, at fixed $p$, the parameter $q$ will govern the matter content and superpotential 
of an $SU(N)^{2p}$ quiver. Recall also 
that at fixed $p$, the central charge $a$ is a monotonic function 
of $q$ which is bounded between the values corresponding to $T^{1,1}/\Z_p$
and $(S^5/\Z_{2})/\Z_p$:
\be
a({T^{1,1}}/\Z_p) < a ({Y^{p,q}}) < a({S^5}/\Z_2\times \Z_p)~,
\ee
suggesting that the different $q$--theories might all be related to 
the same ``parent'' orbifold model. However, we will not pursue this direction 
any further in the present paper. Instead, 
we focus on $Y^{2,1}$ where the dual quiver theory 
is already known. This instance 
already captures many of the essentially new features of these AdS/CFT duals.

\begin{figure}[!th]
\vspace{5mm}
\begin{center}
\epsfig{file=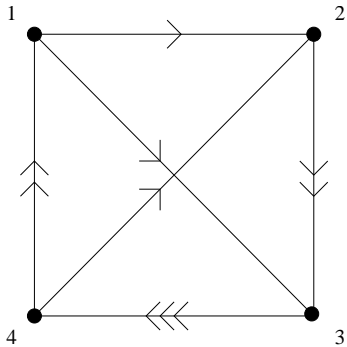,width=4.5cm,height=4.5cm}\\
\end{center}
\caption{Quiver diagram associated to the complex cone over $dP_1$.} 
\label{quiverfig}
\end{figure}

A quiver gauge theory for $dP_1\cong \mathbb{F}_1$ was obtained\footnote{
In this section we denote the first 
del Pezzo surface by $dP_1$.} in 
\cite{hanany} and is presented in Figure \ref{quiverfig}.
Let us briefly recall the notation of these diagrams. The nodes of the diagram represent different 
gauge group factors $U(N)$. Thus the gauge group for the theory is 
$U(N)^4$. An arrow from node $i$ to node $j$ represents a bifundamental 
field in the representation $\overline{\mathbf{N}}\otimes\mathbf{N}$, where the 
first factor denotes the anti--fundamental representation of the $i$th gauge group, 
and the second factor denotes the fundamental representation 
of the $j$th gauge group. We denote these fields as $X_{ij}$. 
Thus the quiver diagram encodes the field 
content of the theory. One must also specify the superpotential. 
This is given by \cite{hanany2}:
\be
W = \epsilon_{\alpha\beta} \mathrm{tr} \left[X^{\alpha}_{34}X^{\beta}_{41}X_{13}\right] - 
\epsilon_{\alpha\beta} \mathrm{tr}\left[X^{\alpha}_{34}X^{\beta}_{23}X_{42}\right] + 
\epsilon_{\alpha\beta} \mathrm{tr}\left[X_{12}X^3_{34}X^{\alpha}_{41}X^{\beta}_{23}\right]
\label{W}\ee
where $\epsilon_{\alpha\beta}\in\{\pm1\}$ and $\alpha,\beta \in \{1,2\}$ are
indices of the non--abelian flavor symmetry group $SU(2)_F$.
Note that each term comes from a closed loop in the quiver. This allows
one to construct gauge--invariant monomials, which may then appear in 
the superpotential. 

One is then particularly interested in the Higgs branch of such a theory. 
This arises by considering $U(N)\rightarrow U(1)^N$ for each gauge 
group factor. One effectively considers the case $N=1$ so that 
the gauge theory is an abelian theory -- the case $N>1$ will simply 
be given by $N$ copies of the $N=1$ case. The fields $X_{ij}$ have 
various charges under the $U(1)^4$ gauge group. Setting the $D$--terms 
of the gauge theory to zero and dividing by the gauge group is, as we 
have discussed already in a different context in this paper, a K\"ahler 
quotient construction, and the result is a toric variety (an overall 
$U(1)$ decouples and is physically 
the  centre of mass $U(1)$ of the D3--branes). 
However, to get the vacuum of the theory one must also set the 
$F$--terms to zero, which means extremising the superpotential: 
$\diff W$=0. This gives a system of relations among the linear 
sigma model fields, which define hypersurfaces in the toric 
variety -- the intersection of these define the Higgs branch of the 
theory, which is part of the moduli space of vacua. 
One can also get to this result by computing all invariant monomials 
in the fields, and then finding all relations among them, including 
those relations given by $\diff W=0$. 
The slightly non--trivial fact is that this is indeed the complex cone over the 
first del Pezzo surface. We will not review this here, but instead 
refer the reader to the literature for details (see e.g. 
\cite{Beasley:1999uz}).
If the quiver gauge theory above is 
interpreted as living on a D3--brane, then this moduli 
space should be the geometry seen by the brane. For $N>1$ one 
has $N$ D3--branes in their Higgs phase, which is why one obtains 
$N$ copies of the above moduli space.

Let us now recall the flavour and R--symmetries of the theory. 
The superpotential above is manifestly invariant
under the non--abelian flavour group $SU(2)_F$, for which the $\alpha,\beta$ indices form a doublet. 
Crucially, there is also a non--anomalous $U(1)\times U(1)$ abelian 
flavour symmetry which is preserved in the IR. 
Taking this into account, the $a$--maximisation calculation applied
to this theory \cite{BB} then gives 
the exact R--charges in the IR. For the sake of clarity, 
these are listed\footnote{We thank the authors of \cite{BB}
for communicating the results of their calculation prior to publication.} in Table \ref{chargesexact}.  
\begin{table}[h!]
\begin{center}
\begin{tabular}{|cc|}
\hline
 $X_{ij}$             & $R_{exact}$ \\
\hline
 $X^{\alpha}_{34}$ & $\tfrac{1}{3}(-1+\sqrt{13})$\\
 $X^{3}_{34}$      & $-3+\sqrt{13}$\\
 $X^{\alpha}_{41}$ & $\tfrac{4}{3}(4-\sqrt{13})$\\
 $X^{\alpha}_{23}$ & $\tfrac{4}{3}(4-\sqrt{13})$\\
 $X_{12}$          & $\tfrac{1}{3}(-17+5\sqrt{13})$\\
 $X_{13}$          & $-3+\sqrt{13}$\\
 $X_{42}$          & $-3+\sqrt{13}$\\
\hline
\end{tabular}
\end{center}
\caption{Exact R--charges computed from $a$--maximisation \cite{BB}.}
\label{chargesexact}
\end{table}
 
Recall that, as proposed in  \cite{IW}, the R--symmetry mixes with the 
abelian flavour symmetries maximising, among all such admissible R--symmetries, 
a certain combination of `t Hooft anomalies. The value of this 
combination of anomalies at the critical point is the exact central charge 
of the theory in the infra--red, and is given by the formula
\be
a=\frac{3}{32}\left(3\mathrm{Tr}R^3-\mathrm{Tr}R\right)~.
\label{defa}
\ee
Substituting the values for the R--charges from Table \ref{chargesexact}
into (\ref{defa}), and comparing with 
(\ref{central}) one finds a corresponding volume
\bea
\frac{13\sqrt{13}+46}{12\cdot 27} \pi^3
\eea
which precisely agrees with the volume of $Y^{2,1}$ (\ref{volume}) on setting 
$p=2$, $q=1$.

Let us finally consider the baryons of the gauge theory.  Recall that 
baryonic operators $B$ of the gauge theory are dual to D3--branes wrapping 
supersymmetric cycles $\Sigma$ in the geometry. Their R--charges  are related to the
volumes of these supersymmetric cycles according to the 
general formula\footnote{We suppress the overall factors of $N$.} \cite{kleb}
\be
R[B] = \frac{2}{3}\cdot \left(\frac{\pi}{2\, \mathrm{vol}(Y)}\right)\cdot 
\mathrm{vol}(\Sigma)~.
\ee
Recall we have shown in Section \ref{section2} that for each manifold $Y^{p,q}$
there are two supersymmetric 3--cycles, which are topologically Lens spaces
$\Sigma_1=S^3/\Z_{p+q}$ and $\Sigma_2=S^3/Z_{p-q}$. We therefore expect that in each case there 
will be two types of baryonic operators $B_1,B_2$ associated to them. 
Substituting for the volume (\ref{volume}) we can write down 
the general formula for the R--charges of the corresponding baryons in the 
$Y^{p,q}$ theory. 
These are given by the unlikely formulae: 
\bea
R[B_1] & = &
\frac{1}{3q^2}\left[-4p^2+2pq+3q^2+(2p-q)\sqrt{4p^2-3q^2} \right]\nn \\
R[B_2] & = & 
\frac{1}{3q^2}\left[-4p^2-2pq+3q^2+(2p+q)\sqrt{4p^2-3q^2} \right]~.
\eea
Note that they are interchanged by changing the sign of $q$. Setting 
$p=2$, $q=1$ the formulae give
\be
R[B_1] = -3+\sqrt{13}, \qquad R[B_2] = \tfrac{1}{3}(-17+5\sqrt{13})~.
\ee
These agree precisely with two of the four different R--charges  
listed in Table \ref{chargesexact}.

\vskip 1cm

\subsection*{Acknowledgments}
\noindent We would like to thank M. Bertolini, F. Bigazzi, A. Hanany,
K. Intriligator, E. Lerman, C. Vafa, D. Waldram, B. Wecht, and  S.--T. Yau  
for discussions and e--mail correspondence. In particular we would like to thank 
E. Lerman for comments on a draft version of this paper.
We are also grateful to the authors of \cite{BB} for earlier collaboration on 
related material, and especially for communicating their $a$--maximisation 
calculation.
DM would like to thank the 2004 Simons Workshop on Mathematics and Physics, 
for hospitality at initial stages of this work.
Part of this work was carried out whilst both authors were postdoctoral 
fellows at Imperial College, London. In particular DM was funded by a 
Marie Curie Individual Fellowship under contract number HPMF-CT-2002-01539,
while JFS was supported by an EPSRC mathematics fellowship. 
At present JFS is supported by
NSF grants DMS--0244464, DMS--0074329 and DMS--9803347.

\appendix

\section{The conifold}

In this appendix we compute the moment cone, gauged linear sigma model and
toric diagram for the conifold, $C(T^{1,1})$. Of course, many of these
results are
well--known in the physics literature -- we include the discussion only as a
simple illustration of the systematic techniques used in this paper, in
the context of an example familiar to many physicists.

The homogeneous Sasaki--Einstein metric on $S^2\times S^3$ is usually
referred to as $T^{1,1}$. The metric is particularly simple \cite{candelas}:
\be
\diff s^2 = \frac{1}{6}(\diff \theta_1^2+ \sin^2\theta_1\diff\phi_1^2
+ \diff \theta_2^2 + \sin^2\theta_2 \diff\phi_2^2) +
\frac{1}{9}(\diff \psi + \cos\theta_1\diff\phi_1 + \cos\theta_2\diff\phi_2)^2~.\ee
Here $\theta_i,\phi_i$, $i=1,2$, are usual polar and axial coordinates
on two round two--spheres, and $\psi$ is a coordinate on a principle
$U(1)$ bundle over $S^2\times S^2$. Here $\psi$ has period $4\pi$ so that
the Chern numbers over the two two--spheres are both equal to 
one\footnote{One may also set $\psi$ to have period $2\pi$ yielding 
$T^{1,1}/\Z_2$ which is a also Sasaki--Einstein {\em manifold}. 
In fact, this is the horizon manifold of the complex cone over $\F_0\simeq \cp^1 \times
\cp^1$. Note that one must be careful to ensure
that the Killing spinors are well--defined on making such identifications.}. 
In particular, $3\de/\de\psi$ is
the Reeb vector so that this is a regular Sasaki--Einstein manifold --
the base K\"ahler--Einstein manifold is just $\cp^1\times\cp^1$.

The symplectic form on the metric cone is
\be
\omega = \frac{1}{6}r^2 \left(\sin\theta_1\diff\theta_1\wedge \diff\phi_1
+ \sin\theta_2\diff\theta_2\wedge\diff\phi_2\right) - \frac{1}{3}r\diff r \wedge
(\diff\psi+\cos\theta_1\diff\phi_1+\cos\theta_2\diff\phi_2)~.
\ee
Clearly we have three commuting Hamiltonian $U(1)$s generated by $\de/\de\phi_i$,
$i=1,2$, and $\de/\de\psi$. As in the main text, one must be careful to
ensure that one picks a basis for an effectively acting $\T^3$ action
when computing the moment map. If one fixes $\theta_1,\phi_1$ on
the first two--sphere, one obtains a copy of $S^3$, written as a principle
$U(1)$ bundle over the second two--sphere. The effectively acting isometry
group on this squashed $S^3$ is $U(2)$, as discussed in the main text.
Defining $2\nu=\psi$, so that $\nu$ has canonical period $2\pi$, one can
therefore take the following basis for the $\T^3$ action:
\bea\label{conbas}
e_1 & = & \frac{\de}{\de\phi_1}+\frac{1}{2}\frac{\de}{\de\nu}\nn\\ \nn
e_2 & = & \frac{\de}{\de\phi_2}+\frac{1}{2}\frac{\de}{\de\nu}\\ 
e_3 & = & \frac{\de}{\de\nu}~.\eea
The corresponding moment map, homogeneous under rescaling of the cone,
is now easily computed to be
\be
\vec{\mu} = \left(\frac{1}{6}r^2(\cos\theta_1+1), \frac{1}{6}r^2
(\cos\theta_2+1), \frac{1}{3}r^2\right)~.\ee
The image of the moment map $\mu:C(T^{1,1})\rightarrow\R^3$ is
a convex rational polyhedral cone generated by the four edge
vectors:
%
%
\bea
\vec{\mu}(NN) & =& \tfrac{1}{3}(1,1,1)\nn\\ \nn
\vec{\mu}(NS) & =& \tfrac{1}{3}(1,0,1)\\ \nn
\vec{\mu}(SN) & =& \tfrac{1}{3}(0,1,1)\\ 
\vec{\mu}(SS) & =& \tfrac{1}{3}(0,0,1)~.
\eea
That is, the subspaces over which a $\T^2$ collapses are precisely
the four subspaces $NN=\{\theta_1=0, \theta_2=0\}$,
$NS=\{\theta_1=0, \theta_2=\pi\}$,  $SN=\{\theta_1=\pi, \theta_2=0\}$,
$SS=\{\theta_1=\pi, \theta_2=\pi\}$
-- these are all copies of the fibre circle over the corresponding point on the base
$S^2\times S^2$. The outward pointing primitive normal vectors
to the cone are computed to be
\be
v_1 = [1,0,-1], \quad v_2 = [0,1,-1], \quad v_3 = [0,-1,0], \quad v_4 =
[-1,0,0]~.\ee
Notice that these indeed form a good cone, as defined in the main
text. Also notice that the vectors $\{v_i\}$ span $\Z^3$ over
$\Z$. Lerman's theorem then states that the base of the
metric cone is simply--connected, which is of course correct.
Moreover, the fact that there are four facets means that $\pi_2(T^{1,1})
\cong \Z$, again correct.

We may now apply the Delzant theorem. The kernel is trivially calculated to 
be
$(1,-1,-1,1)$. Thus the theorem gives a $U(1)$ gauged linear sigma model
on $\C^4$ with charges $(1,-1,-1,1)$ -- this is of course well--known to
give the conifold. Turning on the FI parameter $t>0$, $t<0$ gives the two
small resolutions of the conifold, related by the flop transition.

We now apply the $SL(3;\Z)$ transformation
\be
\left(\begin{array}{ccc}
1 & 1 & 2 \\
0 & -1 & -1 \\
0 & 0 & -1\end{array}\right)\ee
to the torus $\T^3$ of symmetries. The normal vectors now read
\be
v^{\prime}_1 = [-1,1,1], \quad  v^{\prime}_2 = [-1,0,1], \quad
v^{\prime}_3 = [-1,1,0], \quad v^{\prime}_4 = [-1,0,0]~.\ee
These all lie in the plane at $e_1=-1$. Dropping this gives vectors in
$\R^2$:
\be
{\cal V}_1 = [1,1], \quad {\cal V}_2 =[0,1], \quad {\cal V}_3 =[1,0], 
\quad {\cal V}_4 =[0,0]~.
\ee
The toric diagram may thus be embedded in the orbifold $\C^3/\Z_2\times\Z_2$
and is presented below.

\begin{figure}[!th]
\vspace{5mm}
\begin{center}
\epsfig{file=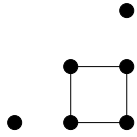,width=1.7cm,height=1.7cm}\\
\end{center}
\caption{Toric diagram of the conifold embedded in the orbifold
$\C^3/\Z_2\times\Z_2$.} \label{con} \vspace{5mm}
\end{figure}
\noindent
We may also analyse the topology of the K\"ahler quotient directly, as in the
main text. The $D$--term constraint reads
\be
|z_1|^2 + |z_2|^2 - |z_3|^2 - |z_4|^2 = t~.
\ee
Setting $t=0$ one obtains a singular space -- the conifold.
Defining gauge invariant coordinates $u=z_1z_3$, $x=z_1z_4$, $y=z_2z_3$, $v=z_2z_4$ we have precisely one relation
$uv=xy$ in $\C^4$, which is thus an equivalent definition of the conifold.

At $z_3=z_4=0$ we have a copy of $\cp^1=S^2$, of size $t$. On a patch in which $z_2\neq0$ we may
introduce a gauge invariant complex coordinate $z=z_1/z_2$. This patch covers a neighbourhood of the south pole at $z_1=0$. Similarly the
coordinate $z^{\prime}=z_2/z_1$ covers a neighbourhood of the north pole at $z_2=0$. Over the intersection of the patches we have the
relation $z^{\prime}=1/z$, thus making the Riemann sphere.
Let us now turn to the remaining coordinates. Consider the subspace in which
$z_2\neq0$ and introduce gauge invariant coordinates $x_2=z_3z_2$, $y_2=z_4z_2$. Thus, over the open set $U_2=\{z_2\neq0, z_3 = z_4 = 0\}
\subset \cp^1$,
our subspace looks like a trivial rank two bundle $\C^2 \times U_2$. Similarly, over $U_1=\{z_1\neq0, z_3 = z_4 = 0\}
\subset \cp^1$ we also have $\C^2 \times U_1$, where the fibre
is coordinatised by
$x_1=z_3z_1$, $y_1=z_4z_1$. On the overlap $U_1\cap U_2$ we have the relation
$x_1=x_2 (z_1/z_2)$, $y_1=y_2 (z_1/z_2)$. By definition, this gluing gives the bundle $\mathcal{O}_{\cp^1}(-1)\oplus\mathcal{O}_{\cp^1}(-1)$, which is
the resolved conifold.

The boundary, or horizon, of this manifold is an $S^3$ bundle over $\cp^1=S^2$, since $S^3$ is the boundary of $\C^2$.
There are various ways of seeing the topology of the horizon.
One way is to
projectivise the original bundle. Recall that to projectivise a rank two complex vector bundle, with transition functions in $U(2)$,
means that one replaces each $\C^2$ fibre with $\cp^1$, and glues the fibres together across overlaps using the
induced transition
functions, which lie in $U(2)/U(1)_D\cong SO(3)$. Here $SO(3)$ acts on the $\cp^1=S^2$ fibre in the usual way. Since the transition functions
of $\mathcal{O}_{\cp^1}(-1)\oplus\mathcal{O}_{\cp^1}(-1)$ are diagonal, the projectivisation is just the product
$\cp^1 \times \cp^1$. The $U(1)$ factor we projected out has unit winding over the fibre $S^2$, since $S^3\rightarrow S^2$ is the Hopf
map which has
Chern number 1. The winding is also $1$ over the base since we began with the sum of two copies of $\mathcal{O}_{\cp^1}(-1)$.
Thus we see explicitly the topology of $T^{1,1}$ as the horizon manifold.

\end{document}